\documentclass[10pt,journal]{IEEEtran}

\ifCLASSOPTIONcompsoc
  \usepackage[nocompress]{cite}
\else
  \usepackage{cite}
\fi

\ifCLASSINFOpdf
\usepackage[pdftex]{graphicx}
\else

\fi

\usepackage{color}
\usepackage[english]{babel}
\usepackage{booktabs}
\usepackage{paralist}
\usepackage[pdftex]{graphicx}
\usepackage[utf8]{inputenc}
\usepackage[T1]{fontenc}
\usepackage{cite}
\usepackage{amsmath,amssymb,amsfonts}
\usepackage{graphicx}
\usepackage{textcomp}
\usepackage{xcolor,soul}
\usepackage{multirow}
\usepackage{balance}
\usepackage{booktabs}
\usepackage[nolist,nohyperlinks]{acronym}
\usepackage{cleveref}
\usepackage{float}
\usepackage{enumerate}
\usepackage{algorithm}
\usepackage[noend]{algpseudocode}
\usepackage{caption}
\usepackage{subcaption}
\usepackage{relsize}
\usepackage{setspace}
\usepackage{url}
\usepackage{texdef2020}

\newcommand{\age}{\Delta}

\usepackage{amstext} 
\usepackage{array}   
\newcolumntype{L}{>{$}l<{$}} 

\usepackage[most]{tcolorbox}

\usepackage{mdframed}
\definecolor{gray90}{gray}{0.90}
\definecolor{gray95}{gray}{0.95}

\newcommand*\rectangled[1]{
\tcbox[enhanced,box align=base,nobeforeafter,colframe=black,size=small]{#1}}

\crefformat{section}{\S#2#1#3} 
\crefformat{subsection}{\S#2#1#3}
\crefformat{subsubsection}{\S#2#1#3}

\newcommand{\RY}[1]{}
\newcommand{\RSP}[1]{}

\newcommand{\agehat}{\hat{\age}}
\newcommand{\RTT}{\text{RTT}}

\renewcommand{\vec}[1]{\begin{bmatrix} #1\end{bmatrix}}

\newcommand{\System}[2]{T_{#1#2}}
\newcommand{\Wait}[2]{W_{#1#2}}
\newcommand{\Service}[2]{S_{#1#2}}

\begin{document}

\bstctlcite{bstctl:nodash}

\title{ACP+: An Age Control Protocol For the Internet}
\author{Tanya~Shreedhar,
        Sanjit~K. Kaul,
        and~Roy~D. Yates
}

\IEEEtitleabstractindextext{%
\begin{abstract}
We present ACP+, an age control protocol, which is a transport layer protocol that regulates the rate at which update packets from a source are sent over the Internet to a monitor. The source would like to keep the average age of sensed information at the monitor to a minimum, given the network conditions. Extensive experimentation helps us shed light on age control over the current Internet and its implications for sources sending updates over a shared wireless access to monitors in the cloud. We also show that many congestion control algorithms proposed over the years for the Transmission Control Protocol (TCP), including hybrid approaches that achieve higher throughputs at lower delays than traditional loss-based congestion control, are unsuitable for age control.

\end{abstract}

}

\maketitle

\IEEEdisplaynontitleabstractindextext

\IEEEpeerreviewmaketitle

\section{Introduction}\label{sec:introduction}

The availability of inexpensive embedded devices with the ability to sense and communicate has led to a proliferation of real-time monitoring applications spanning domains such as health care, smart homes, transportation and natural environment monitoring. 
%
IoT devices are deployed alongside users in homes/offices/cities and connect to the Internet over wireless last-mile access such as WiFi.
Such devices repeatedly sense various physical attributes of a region of interest, for example, traffic flow at an intersection. This results in a device (the \emph{source}) generating a sequence of packets (\emph{updates}) containing measurements of the attributes. A more recently generated update includes a more current measurement. These updates are communicated over the 
network to a remote \emph{monitor} in the cloud, which processes them for analytics and/or to compute any required actuation. 
%

For monitoring applications, it is desirable that \emph{freshly} sensed information is available at monitors.
At a monitor, the freshness of an update  is measured by its {\em age}, the time elapsed since its generation at the source. 
When an application delivers a stream of updates, it is desirable to minimize the time-average age of sensed information at the monitor.
%

%
%
%
While one may achieve a low update packet delay by simply choosing a low rate for the source to send updates, this may be detrimental to freshness;  a low update rate can lead to a large \emph{age} 
at the monitor simply because updates from the source are infrequent.
On the other hand, sending updates at a high rate over the network can also be detrimental to freshness if each received update has high age as a result of experiencing a large network delay.  
In the absence of 
network mechanisms that mitigate congestion, freshness at a monitor is optimized by the source smartly choosing an update rate~\cite{KaulYatesGruteser-Infocom2012,KamKompellaEphremides2013ISIT,KamKompellaEphremides2014ISIT,KamKNE2016IT,yates2020age-survey} 
 appropriately matched to the network.
 
\begin{figure}[!t]             
\begin{center}
\includegraphics[width=0.95\linewidth]{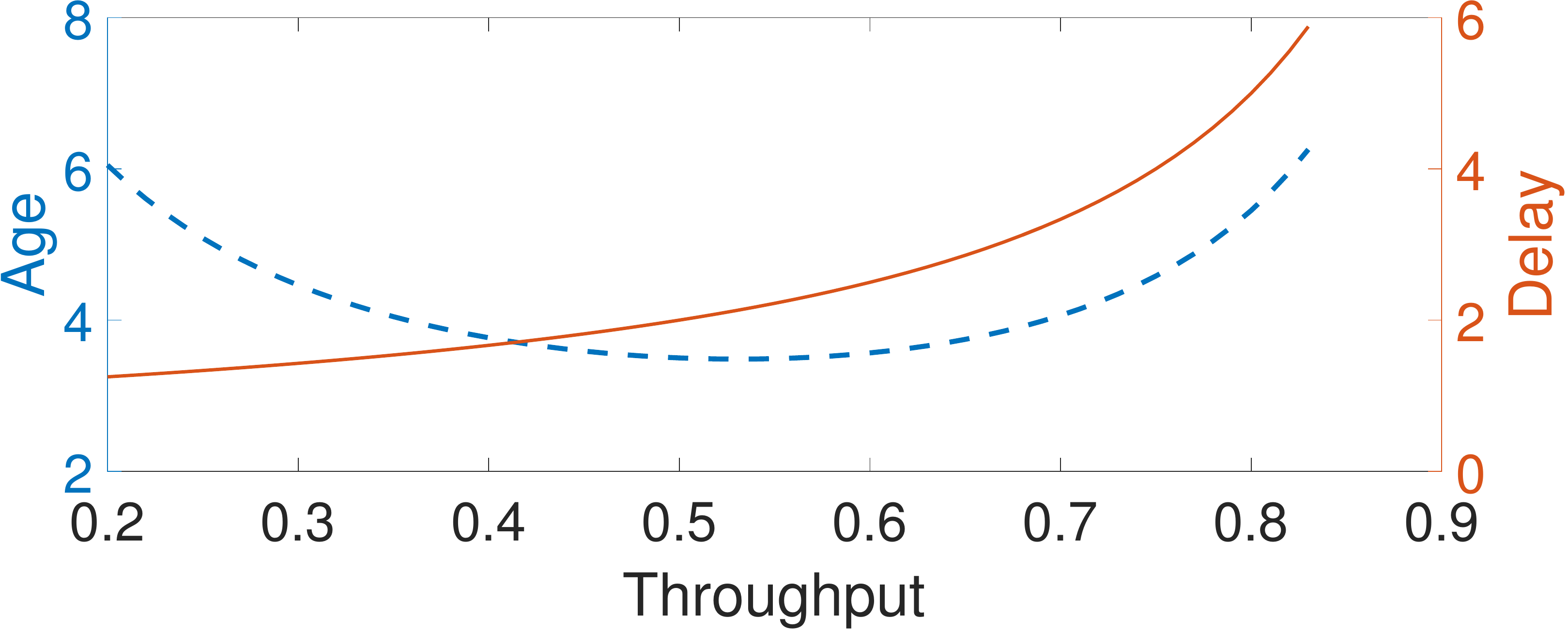}
\caption{Interplay of the networking metrics of delay (solid line), throughput (normalized by service rate) and age. Shown for a M/M/1 queue~\cite{shortle2018fundamentals} with service rate of $1$. The age curve was generated using the analysis for a M/M/1 queue in~\cite{KaulYatesGruteser-Infocom2012}. Under light and moderate loads, throughput (average network utilization) increases linearly in the rate of updates. This leads to an increase in the average packet delay. Large packet delays coincide with large average age. Large age is also seen for small throughputs (small rate of updates). At a low update rate, the monitor receives updates infrequently, and this increases the average age (staleness) of its most fresh update.}
\label{fig:networkingMetrics}
\end{center}
\end{figure}

%
%
Figure~\ref{fig:networkingMetrics} depicts the typical behavior of the metrics of delay and age as a function of throughput, which has been observed for a wide variety of systems that are subject to congestion effects~\cite{yates2020age-survey}. 
Observe that there exists a sending rate (and corresponding throughput) at which age is minimized.


More so than voice/video, monitoring applications are exceptionally loss resilient and they don't benefit from the source retransmitting lost updates. 
This is in contrast to
applications like that of file transfer that require reliable transport and high throughputs but are delay tolerant. Such applications use a variant of Transmission Control Protocol (TCP) for end-to-end delivery of application packets. As we show in Section~\ref{sec:TCP} and later in Section~\ref{sec:ageing_main}, the congestion control algorithm of TCP, which optimizes the use of the network pipe for throughput, and TCP's emphasis on guaranteed and ordered delivery is detrimental to keeping age low. 
Unlike TCP, User Datagram Protocol (UDP) ignores dropped packets and delivers packets to applications as soon as they are received. While this makes it desirable for age-sensitive applications; sending updates at a fixed rate incognizant of the underlying network can be, in fact, disastrous for age of the updates.

In this paper we detail the age control protocol ACP+, which builds on an early version named ACP~\cite{shreedhar_acp_mobicom2018,tanya-kaul-yates-wowmom2019} that was the first proposal for a transport layer age control protocol. ACP+ was first introduced in~\cite{acp_infocom2021} together with the failings of ACP in shared access settings and some initial insights on the age control performance of TCP congestion control algorithms. ACP+ would like to keep the average age of sensed information at the monitor to a minimum.
To achieve the same, ACP+ adapts the rate of updates from a source, in a network transparent manner, to perceived congestion over the end-to-end path between the source and the monitor. Consequently, like ACP, it also limits congestion that would otherwise be introduced by sources sending to their monitors at unnecessarily fast update rates.

Our contributions in this paper include the following.
\begin{enumerate}[(a)]
	\item We investigate the impact of different TCP configurations, such as congestion window and segment sizes, on the age of updates at a monitor, and compare with UDP (User Datagram Protocol). 
	
	\item We detail how ACP+ interfaces with the TCP/IP networking stack via UDP and with sources and monitors. 
	
	\item We define the age control problem over the Internet and intuit a good age control behavior using a mix of analysis and simulations. This leads us to a detailed description of the control algorithm of ACP+. 
	
	\item We provide a detailed evaluation of ACP+ using a mix of simulations (controlled, easier to introduce very high contention, however, only a few hops) and real-world experiments over the Internet (WiFi access with many end-to-end paths sharing it, resulting in low to moderately high contention, followed by many hops over the very fast Internet backhaul).
	
	\item We shed light on age control over end-to-end paths in the current Internet. We observe that the age optimizing rate over an end-to-end path that has a source send updates over a WiFi access followed by the Internet backhaul to a monitor in the cloud is much smaller than the bottleneck link rate of the path, which is the link rate of the WiFi access. The age optimizing rate stays at about $0.5$ Mbps for WiFi access rates of $6$ - $24$ Mbps and backhaul rates as high as $200$ Mbps. In fact, it is the age optimizing rate over the path in the absence of a first WiFi hop. %
    \RY{This is a super interesting observation. Why is the best rate 0.5 Mb/s rather than 0.25 or 1.0 Mb/s? What does it depend on?}\RSP{It is difficult to say why the best rate is a particular value. That it so low is most likely because of the utilization of the internet paths by the other flows. We have added a line saying this below.} %
    Turns out that the intercontinental path, much faster than the WiFi link, is in fact the constraining factor with respect to the achievable age over the end-to-end path, likely because of the other traffic flows that utilize the intercontinental path. %
    \RY{We need to be able to explain this in a sentence or maybe a few sentences.} %
    We also observe that at the age optimal rate, depending on the network scenario, a source may send multiple updates per round-trip-time (RTT) or may send an update over many RTT. In general, the bottleneck link rate and the baseline (updates sent in a stop-and-wait manner) RTT may not shed light on the age optimal rate. 
	
	\item We investigate age, throughput and delay trade-offs obtained when using state-of-the-art TCP congestion control algorithms to transport updates over the Internet. We experiment with a mix of loss-based (Reno~\cite{reno} and CUBIC~\cite{ha2008cubic}), delay-based (Vegas~\cite{brakmo1994tcp}) and hybrid congestion control algorithms (YeAH~\cite{baiocchi2007yeah} and BBR~\cite{cardwell-bbr-2016}) for different settings of receiver buffer size. We conclude that TCP congestion control algorithms are unsuitable for age control. In fact, as contention on the access network increases, the AoI performance of TCP degrades unacceptably. 
\end{enumerate}

The rest of the paper is organized as follows. In the next section, we describe related works. In~\Cref{sec:TCP}, we demonstrate why the mechanisms of TCP are detrimental to minimizing age. In~\Cref{sec:problem}, we define the age control problem.  In~\Cref{sec:acpIntuit}, we use simple queueing models to intuit a good age control protocol and discuss a few challenges.
We detail the Age Control Protocol, how it interfaces with a source and a monitor, and the protocol’s timeline in~\Cref{sec:ACPProtocol}. 
\Cref{sec:algorithm} details the control algorithm that is a part of ACP+. 
This is followed by real-world evaluation over Intercontinental paths and a contended WiFi access in~\Cref{sec:orbit_results}. We discuss simulation setup and results in~\Cref{sec:evaluation_acpNew}. 
We discuss the various congestion control schemes used in the Internet in~\Cref{sec:ageing} and ageing over the Internet using these schemes in~\Cref{sec:ageing_main}.
We conclude in~\Cref{sec:conclusion}.
\section{Related Work}
\label{sec:related}


The queue theoretic analysis using AoI wherein the network and the source(s) are assumed to be described by a service distribution, distribution of arrivals of updates into the queueing facility, any queue management like prioritization and preemption is discussed in~\cite{KaulYatesGruteser-Infocom2012,KamKompellaEphremides2013ISIT,KamKNE2016IT,CostaCodreanuEphremides2014ISIT,HuangModiano2015ISIT,CostaCodreanuEphremides2016,Champati-AG-AOI2018,Inoue-MTT-IT2019,soysal2021age,NajmTelatar-AOI2018,Yates-Kaul-IT2019,Moltafet-TComm2019,Yates-ISIT2018}. 
These works typically carry out analysis that results in the distributional properties of age at the monitor or, more typically, the expected value of the time-average age or the peak age. Such works can help choose an appropriate arrival rate for the one or more sources that are sending packets through the queueing system. The choice of rate is one-shot and doesn't adapt to current network conditions.

There have also been substantial efforts to evaluate and optimize age for multiple sources sharing a communication link
\cite{JiangKrishnmachariZZN-ISIT2018,Yates-Kaul-IT2019,Moltafet-TComm2019}. In particular, near-optimal scheduling based on the Whittle index has been explored in \cite{sun2019closed}. 
There are works on scheduling updates to optimize the age at the monitor~\cite{LuJiLi-Mobihoc2018,bedewy2019minimizing,BedewySunShroff17,kadota-modiano-tmc2021}. 
Notable amongst these are the greedy policy, stationary randomized policy, max-weight policy and Whittle's index policy which uses either the network channel conditions and/or age of the information for making the scheduling decision.
Such works adapt to the network conditions and propose an age-based scheduling policy which is usually \textit{one-shot optimization policy}.

There are works that design a sampling policy as a method to reduce the AoI~\cite{Yates2015ISIT,UpdateorWait-IT2017}. One such approach is the \textit{zero-wait} policy that aims to achieve  maximum throughput and minimum delay but it fails to minimize the AoI especially when the transmission times are heavy tail distributed~\cite{Yates2015ISIT,UpdateorWait-IT2017}. The optimal sampling policy in such cases is a threshold one, either deterministic or randomized.
Sampling policies for unreliable transmissions are considered in ~\cite{arafa2021sample,klugel2019aoi}. 
However, more recently, ~\cite{pan2022optimizing_Infocom} proposes an optimal sampling strategy to optimize data freshness for unreliable transmissions with random forward and backward channels. The proposed policy is based on a randomized threshold strategy where the source waits until the expected estimation error exceeds a threshold before sending a new sample in case of successful transmission. Otherwise, the source sends a new update immediately without waiting. All these policies use optimization theory or optimal stopping rules for minimizing age but  only in the context of a \textit{stop and wait} protocol. 
Our work highlights the need to model multi-hop settings, wherein multiple updates could be queued at any time, to understand optimizing age over modern wide-area IP networks.

\begin{figure}[t]
\centering
\includegraphics[width=0.9\linewidth]{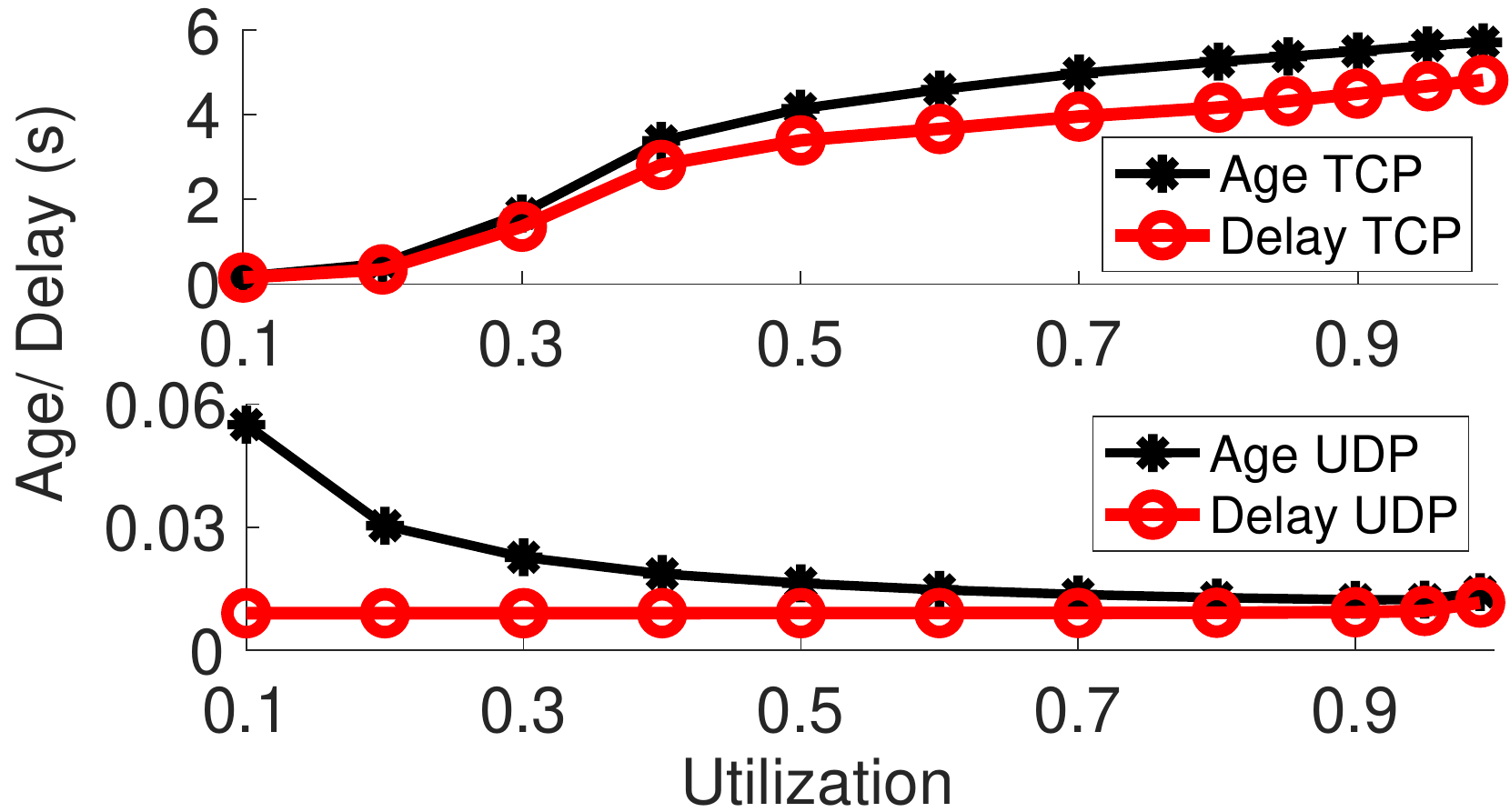}
			\caption{Impact of $0.1$ packet error rate on age and packet delay when using TCP and UDP.}
			\label{fig:UDPvsTCPPacketError}
\end{figure}

While the early work \cite{kaul_minimizing_2011} explored  practical issues such as contention window sizes, the subsequent AoI literature has primarily been focused on analytically tractable simple models. 
Moreover, a model for the system is typically assumed to be known. 
Our objective has been to develop end-to-end updating schemes that perform reasonably well without assuming a particular network configuration or model. This approach attempts to learn (and adapt to time variations in) the condition of the network links from source to monitor. This is similar in spirit to  hybrid ARQ based updating schemes~\cite{Ceran-GG-TWC2019,NajmYatesSoljanin-ISIT2017} that learn the wireless channel. Note that~\cite{Ceran-GG-TWC2019} uses the reinforcement learning techniques in unknown network environments. The chief difference is that hybrid ARQ occurs on the short timescale of a single update delivery while ACP learns what the network supports over many delivered updates.


\noindent \textbf{Age in Systems.}
There is limited systems research on ageing of information and its optimization in real-world networks~\cite{Sonmez-BaghaeeSeaCom2018,shreedhar_acp_mobicom2018,tanya-kaul-yates-wowmom2019,tanya-kaul-yates-arxiv,acp_infocom2021,kadota2020wifresh,acp_esp32,acp_infocom2022}. 
A detailed analysis of all the age related practice works on real networks can be found in~\cite{uysal_practice_2021}.
In \cite{Sonmez-BaghaeeSeaCom2018}, authors discuss the age of information (AoI) in real-networks where a source is sending updates to a monitor over a selection of access networks including WiFi, LTE, 2G/3G and Ethernet. 
The key takeaway from that work is the need for an AoI optimizer that can adapt to changing network topologies and delays. 
Our previous work proposes the Age Control Protocol (ACP)~\cite{shreedhar_acp_mobicom2018,tanya-kaul-yates-wowmom2019,tanya-kaul-yates-arxiv}, which is a transport-layer solution that works in an application-independent and network-transparent manner. ACP attempts to minimize the age of information of a source at a monitor connected via an end-to-end path over the Internet. In~\cite{acp_infocom2021}, we propose a modification to ACP and also compare it with other state-of-the-art TCP congestion control algorithms used in the Internet. In~\cite{kadota2020wifresh}, WiFresh, a MAC and application-layer solution to ageing of updates over a wireless network is proposed.
While both \cite{tanya-kaul-yates-wowmom2019} and \cite{kadota2020wifresh} look at ageing of updates on the Internet, they differ in their approach and scope. ACP is a transport layer solution that works by adapting the source generation rate without any specific knowledge of the access network or any network hop to the monitor, whereas, WiFresh is a scheduling solution designed for WiFi networks. 
In~\cite{acp_infocom2022}, we study the coexistence of age-sensitive ACP+ flows and throughput-hungry TCP flows sharing an end-to-end Internet path over a WiFi access. We found that ACP+ flows coexisting with TCP flows remain unaffected when assigned a higher Differentiated Services Code Point (DSCP) priority when all flows originate in the same device. However, the gains from prioritization vanish when the flows are instead sharing a contended wireless access.

\begin{figure}[t]
\centering
\includegraphics[width=0.9\linewidth]{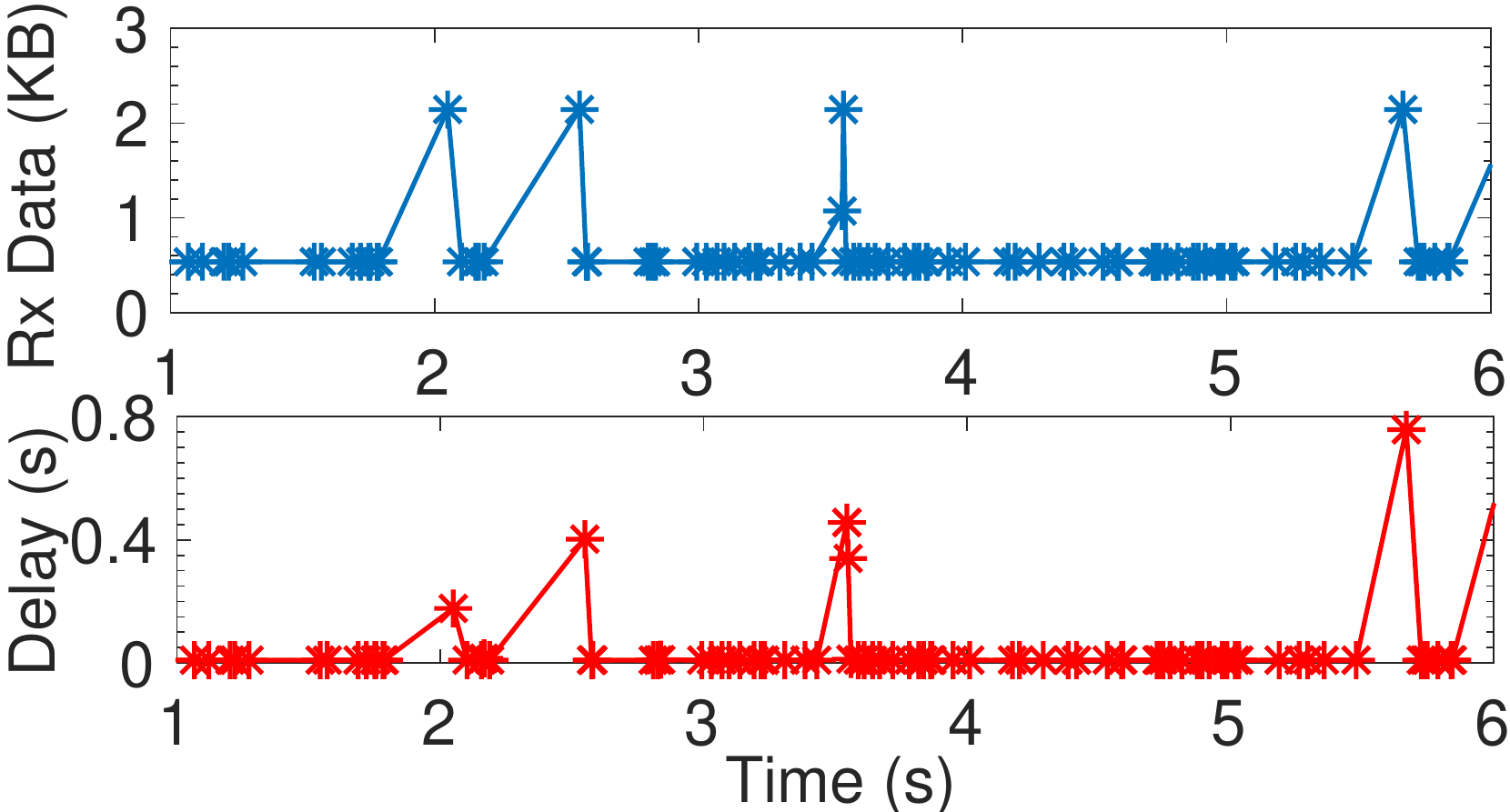}
		\caption{A sample path of delays suffered by update packets transmitted using TCP and the corresponding received bytes by the monitor. The packet error rate was set to $0.1$.}
			\label{fig:UDPvsTCPPacketErrorRXBuffer}
   \label{fig:TCPEval}
  \end{figure} 
  
In summary, unlike other works, ACP and ACP+ aim to provide a practical age control algorithm that seeks to minimize the average age in a multi-source and multi-hop network where there is no prior information about the network and no control on the other sources that have access to it.

\section{Age Sensitive Update Traffic over TCP}
\label{sec:TCP}
\RY{Squeezing three figures on the top of the page made the figs too small. I'm happier to delete some words if we are space constrained. Fig 3 still needs a better caption.  (In Fig 4, why did we compare 500B vs 536B. Is there something special about these sizes?Also how did we vary utilization?}\RSP{The figures have been reorganized, we have added a line to explain how utilization is changed. We have also made clear the choice of $536$ bytes and $500$ bytes.}

Before we delve into the problem of end-to-end age control, we demonstrate why TCP as a choice of transport protocol is unsuitable for age sensitive traffic. Specifically, we show that the congestion control mechanism of TCP, together with its goal of guaranteed and ordered delivery of packets, can lead to  high age at the monitor, particularly in comparison to  UDP. We demonstrate this effect for a wide range of network utilization,
and not just when the utilization is high.

We simulated a simple network consisting of a source that sends measurement updates to a monitor via a single Internet Protocol (IP) router. The source node has a bidirectional point-to-point (P2P) link of rate $1$ Mbps to the router. A similar link connects the router to the monitor. The source uses a TCP client to connect to a TCP server at the monitor and sends its update packets over the resulting TCP connection. We will also compare the obtained age with when UDP is used instead. A larger utilization of the $1$ Mbps links is achieved by the source sending update packets at a faster rate.

\emph{Retransmissions and In-order Delivery:} Figure~\ref{fig:UDPvsTCPPacketError} illustrates the impact of packet errors on TCP. A packet was dropped independently of other packets with probability $0.1$. The update packet size was set to $536$ bytes. The figure compares the average age at the monitor and the average update packet delay, which is the time elapsed between generation of a packet at the source and its delivery at the monitor, when using TCP and UDP. Under TCP, the time-average age achieves a minimum value of $0.18$ seconds when the source utilizes a fraction $0.2$ of the available $1$ Mbps to send update packets. This is much larger than the minimum age of $\approx 0.01$ seconds at a UDP utilization of $\approx 0.8$. 
The large minimum age when using TCP is explained by the way TCP guarantees in order packet delivery to the receiving application (monitor). It causes fresher updates that have arrived out-of-order at the TCP receiver to wait for older updates that have not yet been received, for example, because of packet losses in the network. This can be seen in Figure~\ref{fig:UDPvsTCPPacketErrorRXBuffer} that shows how large measured packet delays coincide with a spike in the number of bytes received by the monitor application. The large delay is that of a received packet that had to undergo a TCP retransmission. The corresponding spike in received bytes, which is preceded by a pause, is because bytes with fresher information received earlier but out of order are held by the TCP receiver till the older packet is received post retransmission. Unlike TCP, UDP ignores dropped packets and delivers packets to applications as soon as they are received. This makes it desirable for age sensitive applications. As we will see later, ACP uses UDP to provide update packets with end-to-end transport over IP~(\cref{sec:ACPProtocol}).

\begin{figure}[t]
    \centering
    \includegraphics[width=0.9\linewidth]{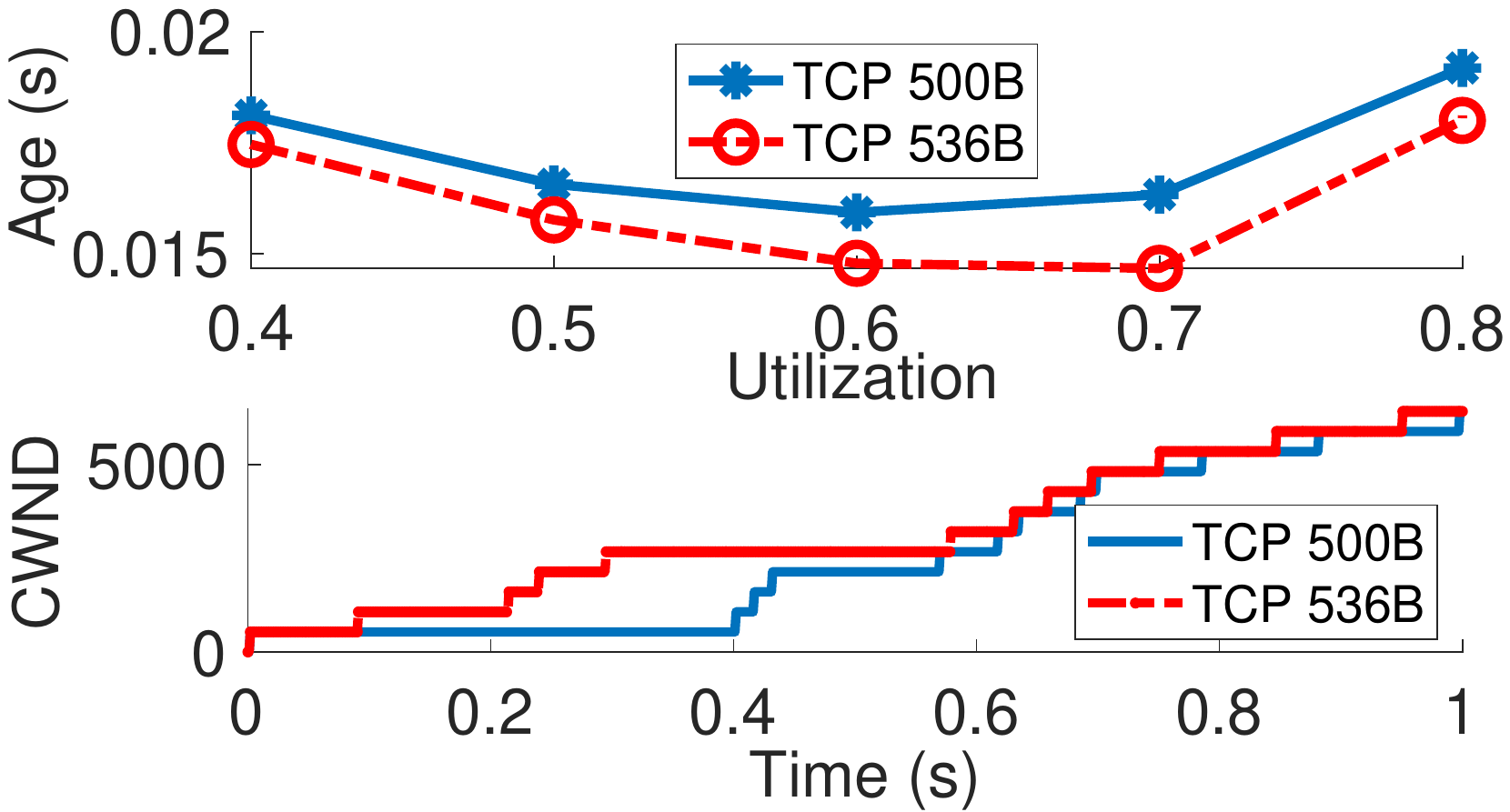}
   \caption{Age as a function of packet size and how packet size impacts increase of the TCP congestion window.}
\label{fig:TCPSmallPacketSizes}
\end{figure}

\emph{TCP Congestion Control and Small Packets:} Next, we describe the impact of small packets, smaller than the minimum sender maximum segment size (SMSS) bytes, on the TCP congestion algorithm and its impact on age. This is especially relevant to a source sending measurement updates as the resulting packets may have small application payloads. Note that no packet errors were introduced in simulations used to make the following observations. The minimum SMSS is $536$ bytes. Observe in the upper plot of Figure~\ref{fig:TCPSmallPacketSizes} that the $500$ byte packet payloads experience higher age at the monitor than the larger $536$ byte packets. The reason is explained by the impact of packet size on how quickly the size of the TCP congestion window (CWND) increases. The congestion window size doesn't increase till SMSS bytes are acknowledged. TCP does this to optimize the overheads associated with sending payload. Packets with fewer bytes may thus require multiple TCP ACK(s) to be received for the congestion window to increase. This explains the slower increase in the size of the congestion window for $500$ byte payloads seen in Figure~\ref{fig:TCPSmallPacketSizes}. This causes smaller packets to wait longer in the TCP send buffer before they are sent out by the TCP sender, which explains the larger age in Figure~\ref{fig:TCPSmallPacketSizes}.
\section{The Age Control Problem}
\label{sec:problem}

\begin{figure}[t] 
	\begin{center}
		\includegraphics[width=0.45\textwidth]{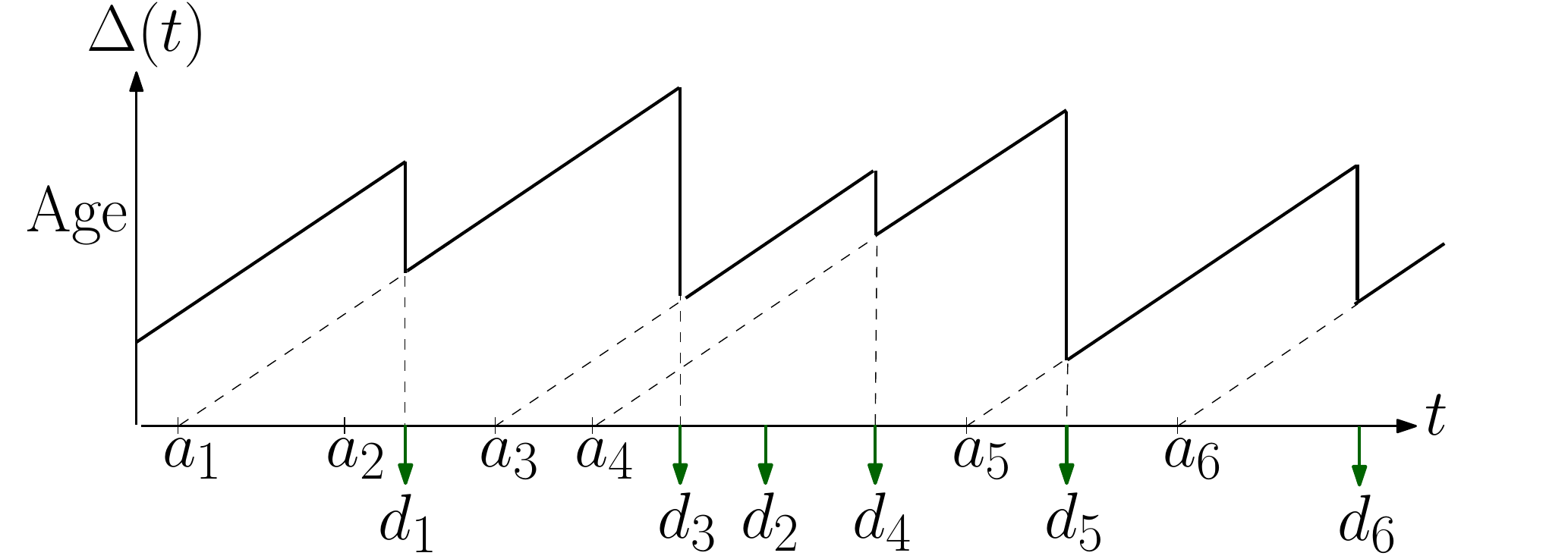}
		\caption{A sample function of the age $\age(t)$. Updates are indexed $1,2,\ldots$. The timestamp of update $i$ is $a_i$. The time at which update $i$ is received by the monitor is $d_i$. Since update $2$ is received out-of-sequence, it doesn't reset the age process.}
		\label{fig:ageSampleFunction}
	\end{center}
\end{figure}

In this section, we formally define the age of sensed information at a monitor. To simplify the presentation, we will assume that the source and monitor are time synchronized, although the functioning of ACP doesn't require the same. Let $z(t)$ be the timestamp of the freshest update received by the monitor up to time $t$. Recall that this is the time the update was generated by the source.

The age at the monitor is $\age(t) = t - z(t)$ of the freshest update available at the monitor at time $t$. An example sample function of the age stochastic process is shown in Figure~\ref{fig:ageSampleFunction}. The figure shows the timestamps $a_1, a_2,\ldots,a_6$ of $6$ packets generated by the source. Packet $i$ is received by the monitor at time $d_i$. At time $d_i$, packet $i$ has age $d_i - a_i$. The age $\age(t)$ at the monitor increases linearly in between reception of updates received in the correct sequence. Specifically, it is reset to the age $d_i - a_i$ of packet $i$, in case packet $i$ is the freshest packet (one with the most recent timestamp) at the monitor at time $d_i$. For example, when update $3$ is received at the monitor, the only other update previously received by the monitor was update $1$. Since update $1$ was generated at time $a_1 < a_3$, the reception of $3$ resets the age to $d_3 - a_3$ at time $d_3$. On the other hand, while update $2$ was sent at a time $a_2 < a_3$, it is delivered out-of-sequence at  time $d_2 > d_3$. So update $2$ is discarded by the monitor ACP and age stays unchanged at time $d_2$.

We want to choose the rate $\lambda$ (updates/second) that minimizes the expected value $\lim_{t \to \infty}E[\age(t)]$ of age at the monitor, where the expectation is over any randomness introduced by the network. Note that in the absence of a priori knowledge of a network model, as is the case with the end-to-end connection over which ACP runs, this expectation is unknown to both source and monitor and must be estimated using measurements. Lastly, we would like to dynamically adapt the rate $\lambda$ to nonstationarities in the network.
\section{Good Age Control Behavior and Challenges}
\label{sec:acpIntuit}

ACP must suggest a rate $\lambda$ updates/second at which a source must send fresh updates to its monitor. ACP must adapt this rate to network conditions. To build intuition, let's suppose that the end-to-end connection is well described by an idealized setting that consists of a single FCFS queue that serves each update in constant time.  An update generated by the source enters the queue, waits for previously queued updates, and then enters service. The monitor receives an update once it completes service. Note that every update must age at least by the (constant) time it spends in service, before it is received by the monitor. It will age more if it ends up waiting for one or more other updates to complete service. 

\begin{figure}[t]             
	\begin{center}
	        \subfloat[Update rate high, delay high, age high]{\includegraphics[width=.4\textwidth]{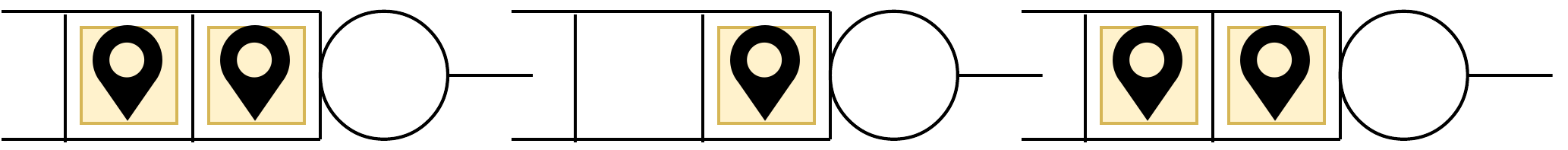}
			\label{fig:goodACP_01}}
			\\
			\subfloat[Update rate low, delay low, age high]{\includegraphics[width=.4\textwidth]{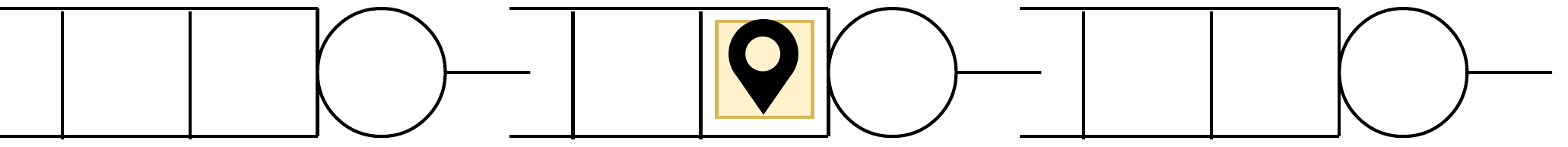}
			\label{fig:goodACP_02}}
			\\
			\subfloat[Ideal snapshot of updates in transit]{\includegraphics[width=.4\textwidth]{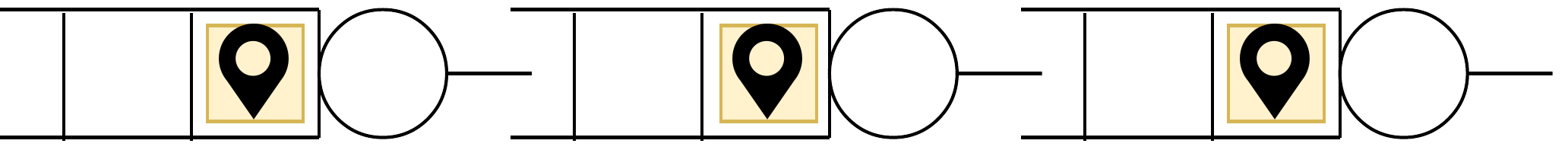}
			\label{fig:goodACP_03}}
		\caption{An illustration of queue occupancy and its impact on age for a network of identical queues with deterministic service in tandem.}
		\label{fig:goodACP}
	\end{center}
\end{figure}

In this idealized setting, one would want a new update to arrive as soon as the last generated update finishes service. To ensure that the age of each update received at the monitor is the minimum, one must choose a rate $\lambda$ such that new updates are generated in a periodic manner with the period set to the time an update spends in service. Also, update generation must be synchronized with service completion instants so that a new update enters the queue as soon as the last update finishes service\footnote{While such just-in-time updating is optimal when the service is deterministic, the same isn't true for general service distributions~\cite{Yates2015ISIT, UpdateorWait-IT2017}.}. In fact, such a rate $\lambda$ is age minimizing even when updates pass through  a sequence of $Q>1$ such queues in tandem. %
\RY{[14] is a textbook (updated Gross and Harris)  and it actually says something about age?? Textbook references need a page number for such a specific observation.} \RSP{The reference was placed incorrectly. Removed it.} %
The update is received by the monitor when it leaves the last queue in the sequence. The rate $\lambda$ will ensure that a generated packet ages exactly $Q$ times the time it spends in the server of any given queue. At any given time, there will be exactly $Q$ update packets in the network, one in each server. Figure~\ref{fig:goodACP} provides an illustration of the above. 

Of course, the assumed network is a gross idealization. We assumed a series of similar constant service time facilities and that the time spent in service and instant of service completion were known exactly. We also assumed lack of any other traffic. However, as we will see further, the resulting intuition is significant. Specifically, \emph{a good age control algorithm must strive to have as many update packets in transit as  possible while simultaneously ensuring that these updates avoid waiting for other previously queued  updates}.

\begin{figure*}[t]             	\begin{center}
		\subfloat[Average Age as a function of \mbox{update} arrival rate $\lambda$]{\includegraphics[width=.31\textwidth]{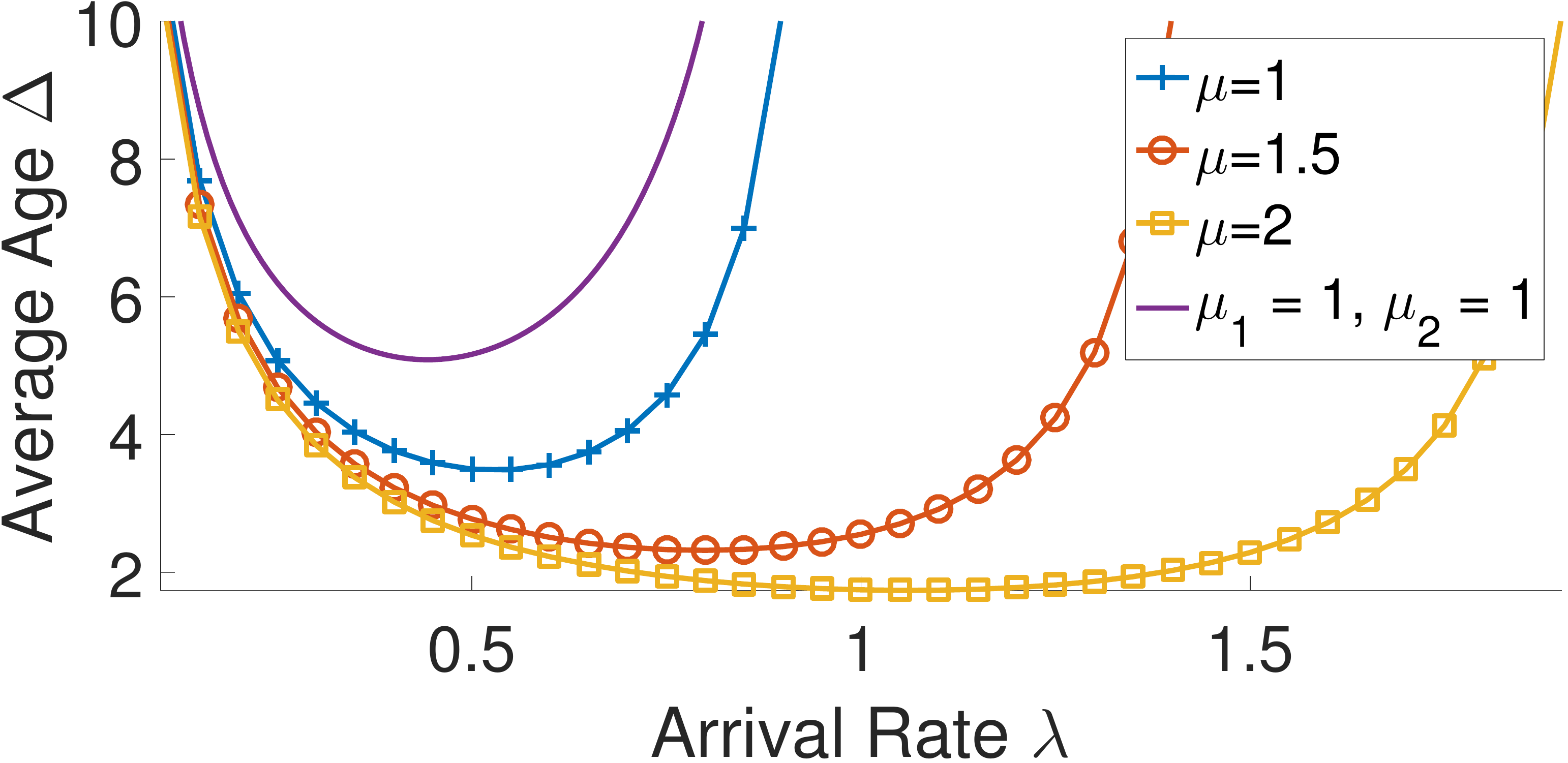}
	\label{fig:motivatingACP_age_lambda}}
		\hspace{0.09in}
		\subfloat[ Average system time as a function of inter-arrival time ($1/\lambda$)]{\includegraphics[width=.31\textwidth]{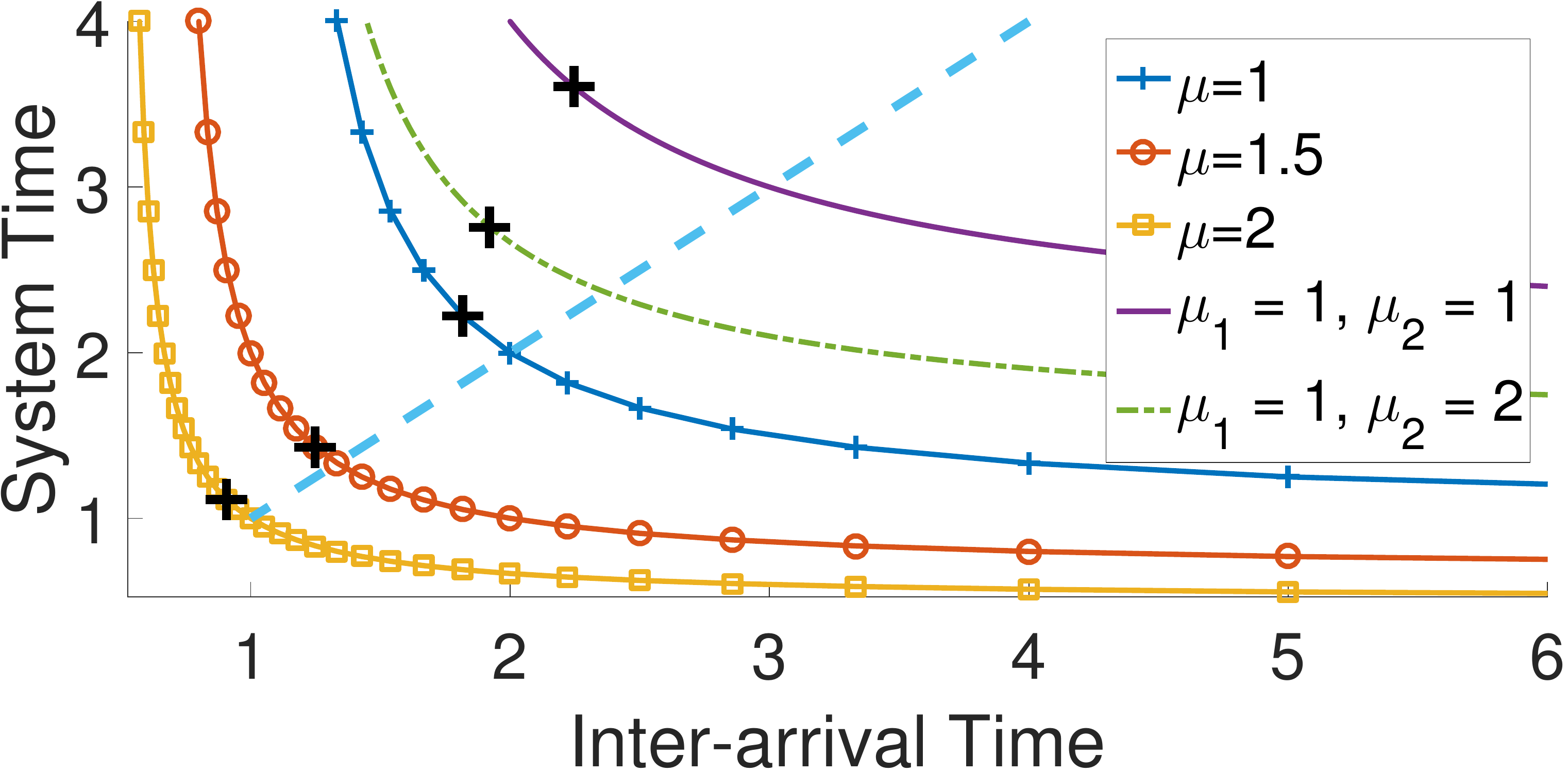}
			\label{fig:motivatingACPArrivalAndSystem}}
		\hspace{0.09in}
		\subfloat[Average backlog as a function of $\mu_2$ with   $\mu_1 = 1$]{\includegraphics[width=.31\textwidth]{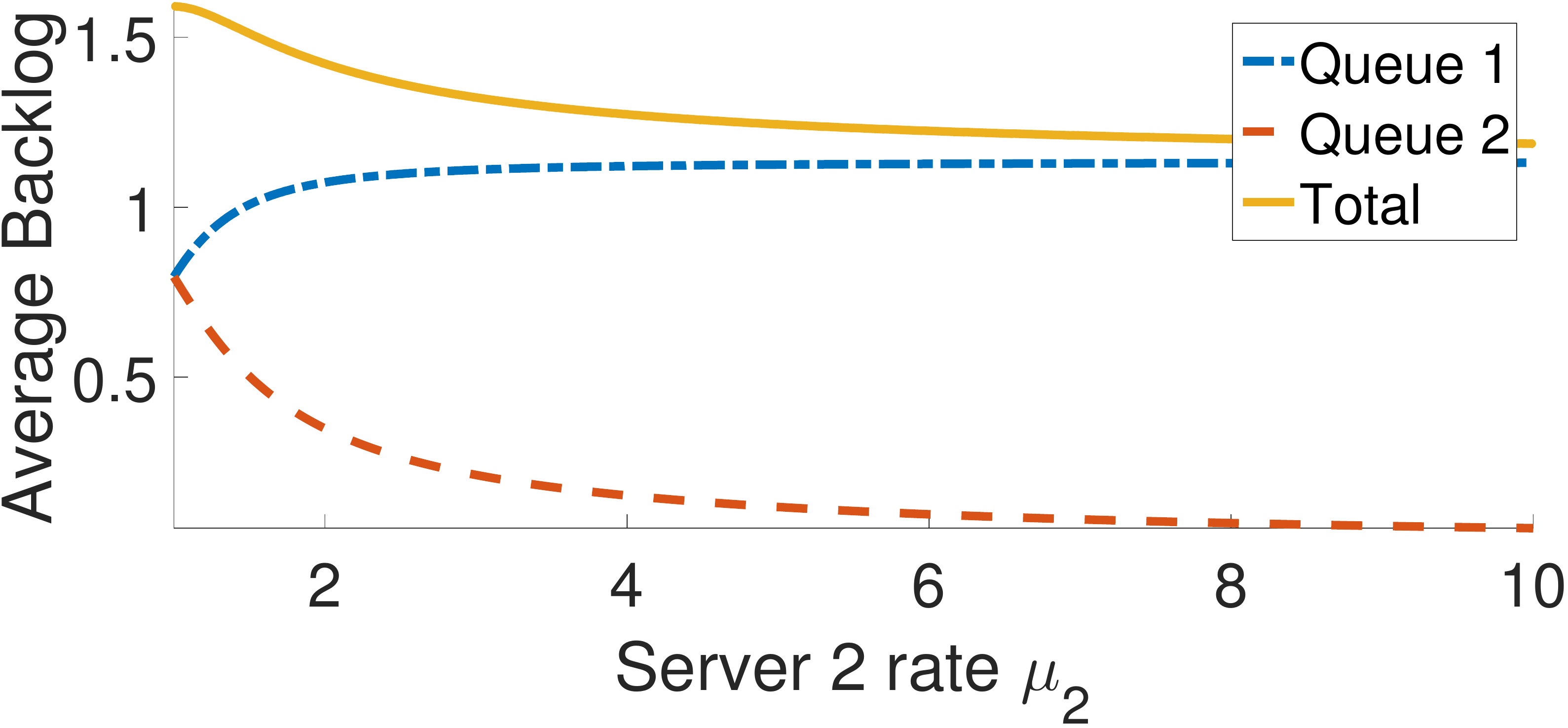}
			\label{fig:motivatingACPBacklog}}
		\caption{Analytical queueing model analysis for two queues having service rates $\mu_1$ and $\mu_2$. }	
		\label{fig:motivatingACP}
	\end{center}
\end{figure*}

Before we detail our proposed control method, we will make a few salient observations using analytical results for simple queueing models and simulation results that capture stochastic service and generation of updates. These will help build on our intuition and also elucidate the challenges of age control over a priori unknown and likely non-stationary end-to-end network conditions. 

\subsection{Analytical Queueing Model for Two Queues} We will consider two queueing models. One is the M/M/1
FCFS queue with an infinite buffer in which a source sends update packets at a rate $\lambda$ to a monitor via a single queue, which services packets at a rate $\mu$ updates per second. The updates are generated as a Poisson process of rate $\lambda$ and packet service times are independent and exponentially distributed with $1/\mu$ as the average service time. In the other model, updates travel through two queues in tandem. Specifically, they enter the first queue that is serviced at the rate $\mu_1$. On finishing service in the first queue, they enter the second queue that services packets at a rate of $\mu_2$. As before, updates arrive to the first queue as a Poisson process and packet service times are exponentially distributed. The average age for the case of a single $M/M/1$ queue was analyzed in~\cite{KaulYatesGruteser-Infocom2012}. We extend their analysis to obtain the analytical expression of average age as a function of $\lambda,\mu_1$ and $\mu_2$ for the two queue case, by using the well known result that updates also enter the second queue as a Poisson process of rate $\lambda$~\cite{shortle2018fundamentals}. The derivation is added as an Appendix. 
\RY{OK so I think the 2 queue analysis is correct. But what do we do about it?? It feels wrong to employ results from an unpublished analysis.} \RSP{Added as an Appendix, which we will submit as part of supplementary material.}

\emph{On the impact of non-stationarity and transient network conditions:} Figure~\ref{fig:motivatingACP_age_lambda} shows the expected value (average) of age as a function of $\lambda$ when the queueing systems are in \emph{steady state}. It is shown for three single $M/M/1$ queues, each with a different service rate, and for two queues in tandem with both servers having the same unit service rate. Observe that all the age curves have a bowl-like shape that captures the fact that a too small or a too large $\lambda$ leads to large age. Such behavior has been observed in non-preemptive queueing disciplines~\cite{yates2020age-survey} in which updates can't preempt other older updates. A reasonable strategy to find the optimal rate thus seems to be one that starts at a certain initial $\lambda$ and changes $\lambda$ in a direction such that a smaller expected age is achieved. 

In practice, the absence of a network model (unknown service distributions and expectations), would require Monte-Carlo estimates of the expected value of age for every choice of $\lambda$. Getting these estimates, however, would require averaging over a large number of instantaneous age samples and would slow down adaptation. This could lead to updates experiencing excessive waiting times when $\lambda$ is too large. Worse, transient network conditions (a run of bad luck) and non-stationarities, for example, because of introduction of other traffic flows, could push these delays to even higher values, leading to an even larger backlog of packets in transit. Figure~\ref{fig:motivatingACP_age_lambda}, illustrates how changes in network conditions (service rate $\mu$ and number of hops (queues)) can lead to large changes in the expected age. 

It is desirable for a good age control algorithm to not allow the end-to-end connection to drift into a high backlog state. 
As we describe in the next section, ACP tracks changes in the average number of backlogged packets and average age over short intervals; when backlog and age increase, ACP acts  rapidly to reduce the backlog.

\emph{On Optimal Average Backlogs:} Figure~\ref{fig:motivatingACPArrivalAndSystem} plots the average packet system time\footnote{The system time of a packet is the time that elapses between its arrival at the first server in the network and its delivery to the monitor.} as a function of the average inter-arrival time ($1/\lambda$) of updates for three single queue $M/M/1$ networks and two networks that have two queues in tandem. As expected, increasing the  inter-arrival time reduces the system time since 
packets wait less for others to complete service. Moreover,  as the inter-arrival time becomes large, the average system time converges to the average service time of a packet. 

For each queueing system, Figure~\ref{fig:motivatingACPArrivalAndSystem} also marks with ($+$) the average inter-arrival time $1/\lambda^*$ that minimizes age. It is instructive to note that for the three single queue systems this inter-arrival time is only slightly smaller than the system time (The blue dashed line in Figure~\ref{fig:motivatingACPArrivalAndSystem} is at $45^\circ$). However, for the two queues in tandem with service rates of $1$ each, the inter-arrival time is much smaller than the system time. The implication being that on an average it is optimal to send slightly more than one ($\approx 1.12$) packet every system time for the single queue system. %
\RY{For M/M/1, my calculation says its 1.12, not 1.2. With $\mu=1$, $\lambda^*=0.53$ and average system time is $T=1/(1-\lambda^*)$ so $\lambda T= 0.53/0.47$? Is that right?}\RSP{Corrected!} %
However, for the two queue network with the similar servers, we want to send a larger number ($\approx 1.6$) of packets every system time. For the two queue network where the second queue is served by a faster server, this number is smaller ($\approx 1.43$). As we observe next, as one of the servers becomes faster, the two queue network becomes more akin to a single queue network with the slower server.

\begin{table}[t]
\begin{center}
\begin{tabular}{@{}ccccccc@{}}
\toprule
Network & $R_1$ & $R_2$ & $R_3$ & $R_4$ & $R_5$ & $R_6$ \\ \midrule
Net A   & 1     & 1     & 1     & 1     & 1     & 1     \\
Net B   & 1     & 1     & 5     & 5     & 1     & 1     \\
Net C   & 1     & 5     & 5     & 5     & 5     & 1     \\
Net D   & 5     & 5     & 5     & 5     & 5     & 1     \\
Net E   & 5     & 5     & 5     & 5     & 5     & 5     \\ \bottomrule
\end{tabular}
\caption{Various P2P link configurations applied to the six-hop network diagram in Figure~\ref{fig:simulationNetwork}. The rates $R_i$ are in Mbps. $R_1$ is the rate of the link between the source and AP-1 and $R_6$ is that of the link between AP-2 and the monitor.}
		\label{tab:goodACPSimulationTable}
		\end{center}
\end{table}

Figure~\ref{fig:motivatingACPBacklog} shows how the optimal average backlog varies as a function of service rate(s). For each choice of service rate (i.e., $\mu$ in the single queue or $\mu_1,\mu_2$ in the tandem queue), the age-optimal arrival rate $\lambda^*$ is selected.  For fixed $\mu_1=1$, 
$\lambda^*$ increases as $\mu_2$ increases, and this causes the average backlog  in queue $1$ to increase. However, as 
$\lambda^*$ increases more slowly than $\mu_2$, the backlog in queue $2$ decreases,  while the sum backlog approaches the optimal backlog for the single queue system. Specifically, as queue $1$ becomes a larger rate bottleneck relative to queue $2$, the optimal $\lambda^*$ must adapt to the bottleneck queue. 

These observations stay the same on swapping $\mu_1$ and $\mu_2$. %
\RY{Is this really true?? I meant to recheck this.} \RSP{Yes, it is! A rewrite (now done) of the expression for AoI makes this fact explicit.} %
Moreover, when the rates $\mu_1$ and $\mu_2$ are similar, the queues see similar backlogs. Notably, when $\mu_1=\mu_2 = 1$, the backlog per queue is smaller than in a network with only a single such queue. However, the sum backlog ($\approx 1.6$) is larger. 

\subsection{Simulating Larger Number of Hops} To see if this intuition generalizes beyond two hops, we simulated an end-to-end connection which has the source send its packet to the monitor over $6$ hops, where each hop is serviced by a bidirectional P2P link\footnote{The hops are illustrated in Figure~\ref{fig:simulationNetwork} in Section~\ref{sec:evaluation_acpNew} on simulations.}. We vary the rates at which the P2P links transmit packets to gain insight into how queues in a network must be populated with update packets at an age optimal rate. We also introduce other traffic in the network that occupies, on an average, $0.2$ Mbps of each P2P link from the source to the monitor. The different configurations are summarized in Table~\ref{tab:goodACPSimulationTable}. For each network configuration we have the source send updates over UDP to the monitor using an a priori chosen rate $\lambda$. We vary $\lambda$ over a range of values and for each $\lambda$ we calculate the obtained time-average age. We empirically identify the age minimizing $\lambda$ for each given network. 

\begin{figure}[!t]             
	\begin{center}
			\includegraphics[width=.45\textwidth]{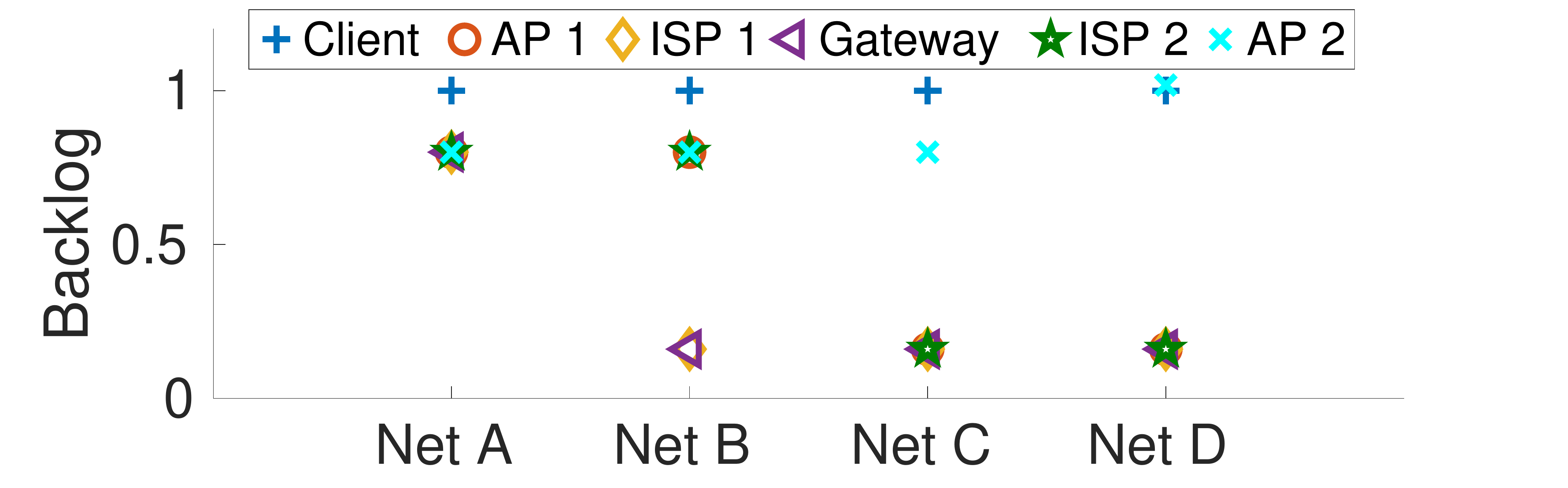}
			\caption{Average packet backlogs at different nodes in the network of Fig~\ref{fig:simulationNetwork}, at the optimal update rate. Net E is similar to Net A and not shown.}
			\label{fig:AverageBacklog_nodes}
	\end{center}
\end{figure}

Figure~\ref{fig:AverageBacklog_nodes} shows the time-average \emph{backlog} (queue occupancy) at the different nodes in the network at the optimal $\lambda$. The backlog at a node includes the update packet being transmitted on a node's outgoing P2P link and any update that is awaiting transmission at the node. Observe that all P2P links in each of Net A and Net E have the same rate, $1$ and $5$ Mbps respectively. Though Net B has links much faster than that of Net A, for both these networks the average backlog at all nodes is close to $1$. That it is smaller than $1$ is explained by the presence of the other flow. The other flow, which also originates at the client, is also the reason why the client sees a slightly larger average queue backlog.

Net B has faster P2P links connecting ISP(s) and the Gateway when compared to Net A. However, its other links are slower than that in Net E. We see that the nodes that have fast outgoing links have low backlogs and those that have slow links have an average backlog close to $0.8$. The source has a slow outgoing link and as a result of the other flow sees slightly larger occupancy of update packets. In Figure~\ref{fig:AverageBacklog_nodes}, similar observations apply to for Net~C and Net~D.  In summary, at the age-minimizing update rate $\lambda$, \emph{links that are relative bottlenecks each see an average backlog of no more than one update packet}. Naturally, nodes with faster links see smaller backlogs in proportion to how fast their links are relative to the bottleneck.

A corollary to the above observations 
is that a good age control algorithm should on an average have a larger number of packets simultaneously in transit in a network with a larger number of hops (nodes/queues). \RY{Or maybe a good age control algorithm will have about as many (or maybe no more)  packets simultaneously in transit as hops?}\RSP{It is not clear if we shouldn't have more than one per hop or only have close to $1$ for any setting. It may be more generally true, however, that one must have more updates over a path that has more hops.}
\section{The Age Control Protocol}
\label{sec:ACPProtocol}

\begin{figure}[!t]             
	\begin{center}
		\includegraphics[width=0.45\textwidth]{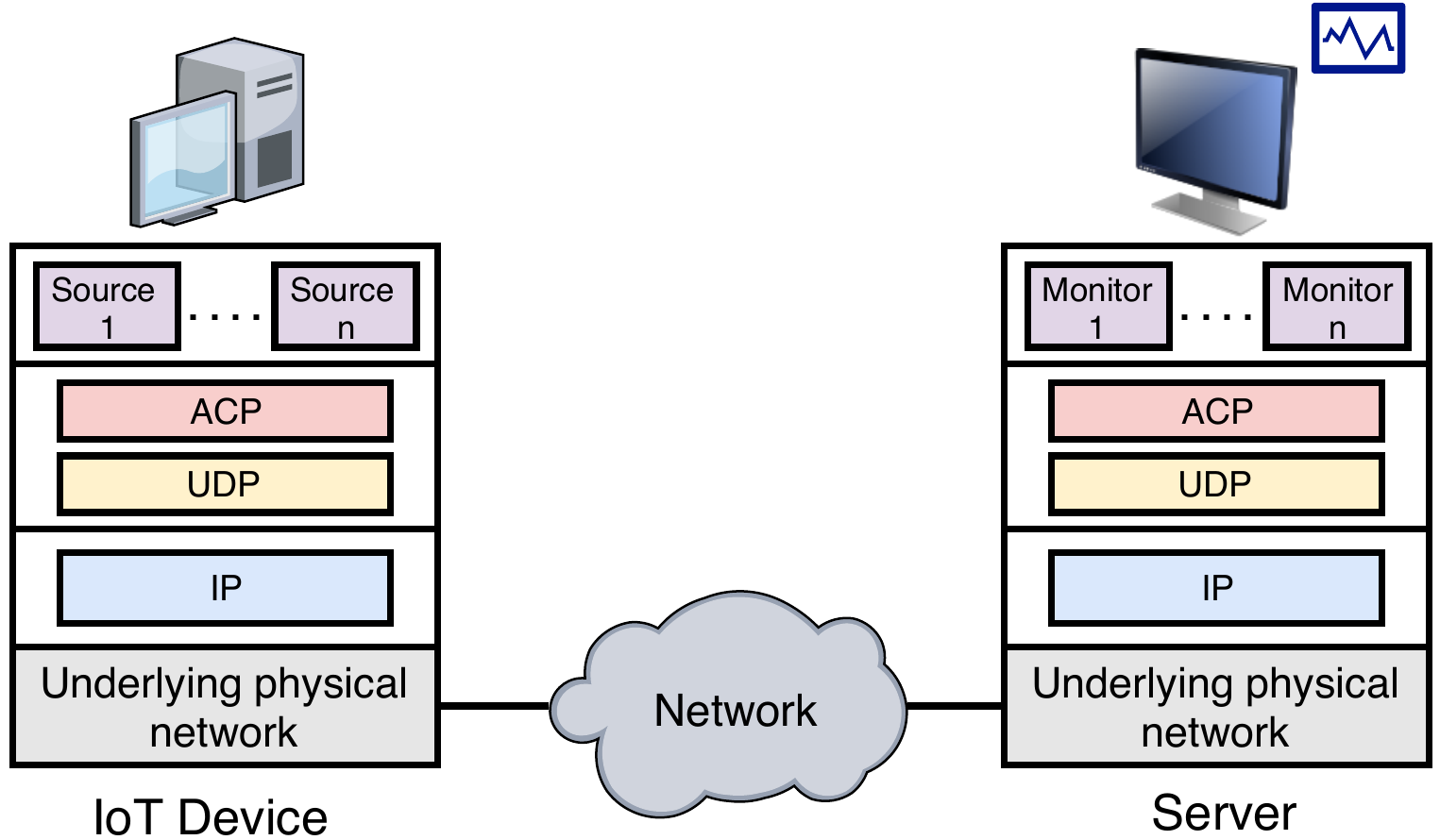}
		\caption{The ACP end-to-end connection.}
		\label{fig:networkStack}
	\end{center}
\end{figure}

The Age Control Protocol resides in the transport layer of the TCP/IP networking stack and operates only on the end hosts. Figure~\ref{fig:networkStack} shows an \emph{end-to-end connection} between two hosts, an IoT device, and a server, over the Internet. A source opens an ACP\footnote{We use ACP when describing aspects common to ACP and ACP+.} connection to its monitor. Multiple sources may connect to the same monitor. ACP uses the unreliable transport provided by the user datagram protocol (UDP) for sending of updates generated by the sources. 


The source ACP appends a header to an update from a source. The header contains a \emph{timestamp} field that stores the time the update was generated. The source ACP suggests to the source the rate at which it should generate updates. To be able to calculate the suggested rate, the source ACP must estimate network conditions over the end-to-end path to the monitor ACP. This is achieved by having the monitor ACP acknowledge each update packet received from the source ACP by sending an ACK packet in return. The ACK contains the timestamp of the update being acknowledged. The ACK(s) allow the source ACP to keep an estimate of the age of sensed information at the monitor. 
An \emph{out-of-sequence} ACK, which is an ACK received after an ACK corresponding to a more recent update packet, is discarded by the source ACP. Similarly, an update that is received \emph{out-of-sequence} is discarded by the monitor. This is because the monitor has already received a more recent measurement from the source.

\begin{figure}[!t]             
	\begin{center}
		\includegraphics[width=0.45\textwidth]{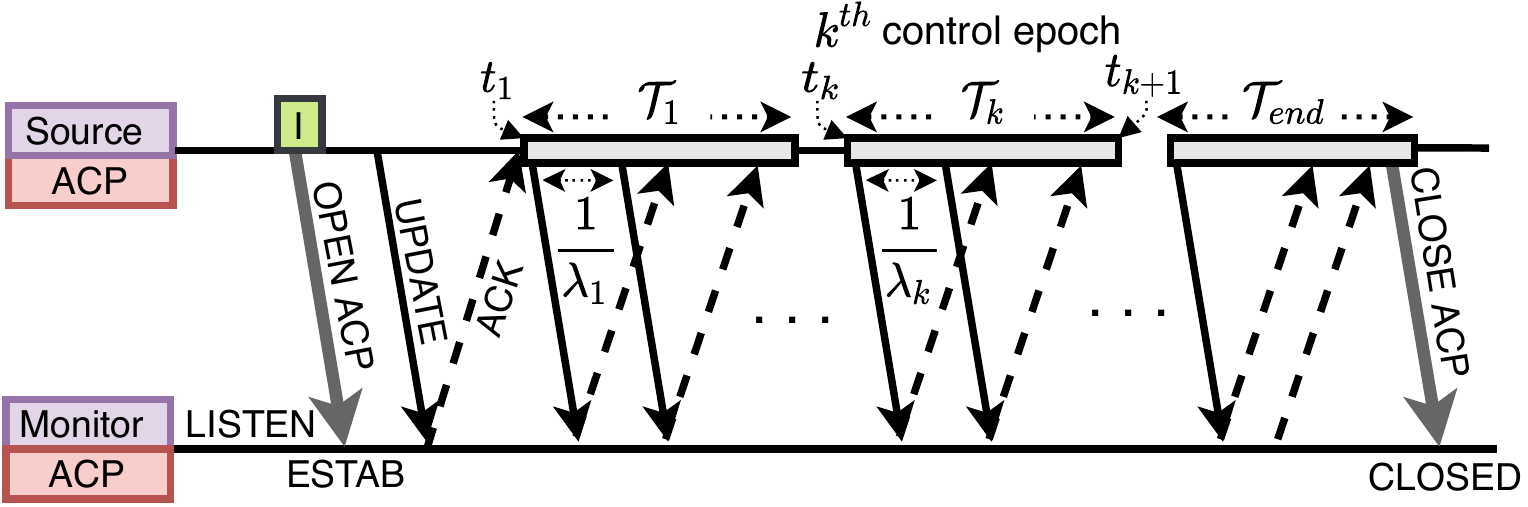}
		\caption{Timeline of an ACP connection.\rectangled{I} marks the beginning of the initialization phase. The control algorithm (Algorithm~\ref{alg:acp+}) is executed when a new control epoch begins.}
		\label{fig:acpConnectionTimeline}
	\end{center}
\end{figure}

Figure~\ref{fig:acpConnectionTimeline} shows a timeline of a typical ACP connection. For an ACP connection to take place, the monitor ACP must be listening on a previously advertised UDP port. The ACP source first establishes a UDP connection with the monitor. This is followed by an \emph{initialization} phase during which the source sends an update and waits for an ACK or for a suitable timeout to occur, and repeats this process for a few times, with the goal of probing the network to set an initial update rate. Following this phase, the ACP connection may be described by a sequence of \emph{control epochs}. The end of the \emph{initialization} phase marks the start of the first control epoch. At the beginning of each control epoch, ACP sets the rate at which updates generated from the source are sent until the beginning of the next epoch. 
We detail our algorithm in the following section.

\section{The ACP+ Control Algorithm}
\label{sec:algorithm}

Let the control epochs of ACP+ be indexed $1,2,\ldots$ such that  epoch $k$ starts at time $t_k$ and ends at $t_{k+1}$, which is the beginning of the next epoch. Let $\lambda_k$ be the update rate set by the control algorithm for epoch $k$. The source ACP transmits updates at a fixed time period of $1/\lambda_{k}$ seconds in the interval $(t_{k}, t_{k+1})$. For the first control epoch, beginning at $t_1$, the update rate $\lambda_{1}$ is set to the inverse of the average packet round-trip-times (RTT) obtained at the end of the initialization phase. 

The length $\mathcal{T}_k=t_{k+1}-t_k$ of the $k\textsuperscript{th}$ control epoch is set as an integral number $\eta$ of time periods, each of length $1/\lambda_k$. That is $\mathcal{T}_k= \eta/\lambda_k$. A control epoch should be long enough to allow for reasonable estimates of the time-average age and backlog that result from setting the update rate to $\lambda_k$ for epoch $k$. In this work, we find $\eta=10$ to work well for all our chosen scenarios in simulations and real-world experiments. 

An ACP source sends updates indexed $i=1,2,\ldots$ with the  $i$\textsuperscript{th} update having timestamp $a_i$. Let $d_i$ be the time the update is received by the monitor and $\dhat_i$ be the time the ACK corresponding to it is received by the source. 

At the source, the ACKs received in sequence are used to construct approximations $\agehat(t)$ and $\Bhat(t)$ of the age process $\age(t)$ at the monitor and the backlog process $B(t)$ that tracks the number of update packets currently in the network on their way to the monitor. To estimate $B(t)$, we define 
\begin{subequations}
\begin{align}
S(t)&=\max\set{i\colon a_i\le t},\\
\Nhat(t)&=\max\{i\colon \dhat_i\le t\}. 
\end{align}
\end{subequations}
At time $t$, $S(t)$ is the index of the freshest update that has been transmitted and $\Nhat(t)$
is the index of the freshest update for which an in-sequence ACK has been received at the source. The estimated backlog at time $t$ is
\begin{equation}
    \Bhat(t)=S(t)-\Nhat(t).\eqnlabel{estimatedB}
\end{equation}
In \eqnref{estimatedB}, the estimated backlog increases by $1$ when the source sends a new update at time $t=a_i$. Further, out-of-sequence ACKs, which the source discards, won't change $\Nhat(t)$ and won't reduce the estimated backlog. Also, when the ACK corresponding to an update $i$ is received in sequence, update $i$ as well as any unacknowledged updates older than $i$ are removed from the estimated backlog. This estimated backlog undercounts updates that arrive out-of-sequence at the monitor. In the absence of out-of-sequence updates, it is a conservative estimate in that updates are removed from the backlog only when ACKs are returned to the source.

To estimate the age process at the source, we define
\begin{equation}\eqnlabel{agehat}
    \agehat(t)=t- a_{\Nhat(t)}.
\end{equation}
The effect of \eqnref{agehat} is that only in-sequence ACKs reduce the estimated age $\agehat(t)$. Moreover, if the ACK for update $i$ is received in sequence, then $\agehat(\dhat_i)=\dhat_i - a_i$, which is the RTT corresponding to update $i$. Resetting the estimate of age to the RTT overestimates the true age $d_i - a_i$ of the update packet, when it was received at the monitor, by the time taken to send the ACK from the monitor to the source. However, this estimate is useful in practice as it doesn't require any time synchronization between the ACP source and monitor. Further, the overestimation offsets the time-average age by the one-way delay from the monitor to the source, suffered by the ACK, and doesn't affect the update rate that minimizes the average age. This has also been shown in~\cite{uysal_practice_2021}.

The ACP source also maintains exponentially weighted moving averages (EWMAs) $\overline{\RTT}$ and $\overline{Z}$ of the RTT and the ACK interarrival time. When an ACK is received in sequence, ACP calculates its RTT and the time elapsed since the last received in-sequence ACK to update the corresponding averages.
When an ACK of update $i$ is received in sequence, the instantaneous value of RTT is set as
\begin{equation}
\RTT_{i}=\dhat_i-a_i
\end{equation}
and that of ACK inter-arrival time is set as
\begin{equation}
 Z_i=\dhat_{\Nhat(\dhat_i)}-\dhat_{\Nhat(\dhat_i^-)}.
\end{equation}
Here $\dhat_i^-$ is the time just before update $i$ is received and $\Nhat(\dhat_i^-)$ is the index of the update whose in-sequence ACK preceded the currently received ACK. The EWMAs $\overline{\RTT}$ and $\overline{Z}$ are updated at $t = \dhat_i$ as
\begin{subequations}
\begin{align}
    \overline{\RTT}(t) &= (1 - \alpha) \overline{\RTT}(t^-) + \alpha \RTT_i,\\
    \overline{Z}(t) &=(1-\alpha)\overline{Z}(t^-)+\alpha Z_i.
\end{align}
\end{subequations}
In the absence of an in-sequence ACK, $\overline{\RTT}(t)$ and $\overline{Z}(t)$ remain unchanged.

Let $\overline{\age}_k$ be the estimate at the  ACP+ source at time $t_k$ of the time-average update age at the monitor. It is obtained by calculating the area under the estimated age function $\agehat(t)$ over $(t_{k-1}, t_k)$ and dividing it by $t_k - t_{k-1}$. Similarly, define $\overline{B}_k$ to be the time-average of backlog $\Bhat(t)$ calculated over the interval $(t_{k-1}, t_k)$.

At time $t_k$, the source ACP+ calculates the difference $\delta_k = \overline{\age}_k - \overline{\age}_{k-1}$ in average age measured over epochs $(t_{k-1}, t_k)$ and $(t_{k-1}, t_{k-2})$. Similarly, it calculates $b_k = \overline{B}_k - \overline{B}_{k-1}$.
ACP+ at the source chooses an action $u_k$ at the beginning of the $k$\textsuperscript{th} epoch that targets a change $b^{*}_{k+1}$ in average backlog over the interval of length $\mathcal{T}_k$ of the $k$\textsuperscript{th} epoch.

\begin{algorithm}[t]
\caption{ACP+ Control Algorithm}
\label{alg:acp+}
\begin{algorithmic}[1]
\State \textbf{INPUT:} $b_k$, $\delta_k$ 
\State \textbf{INIT:} $\text{flag} \gets 0$, $\gamma \gets 0$
\While{true} 
	\If {$b_k>0$  \&\& $\delta_k >0$}\label{alg:one}
		 \If {$\text{flag}==1$}
		 \State $\gamma=\gamma+1$\label{alg:oneincr}
			\State MDEC($\gamma$): $b^{*}_{k+1} = -(1 - 2^{-\gamma}) B_k$
		\Else
			\State DEC: $b^{*}_{k+1} = -1$ \label{alg:oneDEC}
		\EndIf
		\State $\text{flag}\gets 1$
	\ElsIf { $b_k>0$  \&\& $\delta_k <0$}\label{alg:two}
	        \State INC: $b^{*}_{k+1} = 1$
			\State $\text{flag}\gets 0$,
			$\gamma\gets0$ \label{alg:two2}
	\ElsIf { $b_k<0$  \&\& $\delta_k >0$}\label{alg:three}
	    \State INC: $b^{*}_{k+1} = 1$
		\State $\text{flag}\gets 0$, $\gamma\gets0$ 
	\ElsIf { $b_k<0$  \&\& $\delta_k <0$}\label{alg:four}
		 \If {$\text{flag}==1$ \&\& $\gamma>0$}
			\State MDEC($\gamma$): $b^{*}_{k+1} = -(1 - 2^{-\gamma}) B_k$ \label{alg:fourMDEC}
		\Else
		    \State DEC: $b^{*}_{k+1} = -1$
			\State $\text{flag}\gets 0$, $\gamma\gets0$
		\EndIf
	\EndIf
\State \Call{UpdateLambda}{$b^{*}_{k+1}$} \label{alg:update}
\State wait $\overline{T}$ 	
\EndWhile
\Statex
\Function{UpdateLambda}{$b^{*}_{k+1}$}
\State $\lambda_k = \frac{1}{\overline{Z}} + \frac{b^{*}_{k+1}}{\mathcal{\overline{\RTT}}}$ \label{alg:calc}
	\If {$\lambda_k < 0.75*\lambda_{k-1} $}\label{alg:min}
		 \State 
		 $\lambda_k = 0.75*\lambda_{k-1}$
		 \Comment{Minimum $\lambda$ threshold}
	\ElsIf {$\lambda_k > 1.25*\lambda_{k-1} $}\label{alg:max}
	        \State
	        $\lambda_k = 1.25*\lambda_{k-1} $
			\Comment{Maximum $\lambda$ threshold}
	\EndIf
\State
\Return $\lambda_k$
\EndFunction
\end{algorithmic}
\end{algorithm}

The actions, may be broadly classified into additive increase (INC), additive decrease (DEC), and multiplicative decrease (MDEC). MDEC corresponds to a set of actions $\set{\text{MDEC}(\gamma)\colon \gamma=1,2,\ldots}$. Specifically,
\begin{subequations}
\begin{align}
\text{INC}\colon&  b^{*}_{k+1} = 1,\\
\text{DEC}\colon&  b^{*}_{k+1} = -1,\\
\text{MDEC}(\gamma)\colon&  b^{*}_{k+1} = -(1 - 2^{-\gamma}) B_k.
\end{align}
\end{subequations}

ACP+ attempts to achieve $b^{*}_{k+1}$ by setting $\lambda_k$ appropriately. The estimate of $\overline{Z}$ at the source ACP+ of the average inter-update arrival time at the monitor gives us the rate $1/\overline{Z}$ at which updates sent by the source currently arrive at the monitor. The change in the average backlog that will result from setting the rate to $\lambda_k$ is $(\lambda_k - (1/\overline{Z})) \mathcal{\overline{\RTT}}$.  This and $\lambda_k$ allow us to estimate the average change in backlog over $\mathcal{T}_k$ as $(\lambda_k - (1/\overline{Z})) \mathcal{\overline{\RTT}}$. 
Therefore, a desired change of $b^{*}_{k+1}$ requires choosing 
\begin{align}
\lambda_k = \frac{1}{\overline{Z}} + \frac{b^{*}_{k+1}}{\mathcal{\overline{\RTT}}}.
\end{align}

Algorithm~\ref{alg:acp+} summarizes how ACP+ chooses its action $u_k$ as a function of $b_k$ and $\delta_k$. The source ACP+ targets a reduction in average backlog over epoch $k$ in case either $b_k,\delta_k>0$, 
or $b_k,\delta_k<0$.
Both $b_k$ and $\delta_k$ being positive (line~\ref{alg:one} in Algorithm~1) indicates that the update rate is such that updates are experiencing larger than optimal delays. In this case, ACP+ attempts to reduce the backlog, first using DEC (line~\ref{alg:oneDEC}), followed by multiplicative reduction MDEC to reduce congestion delays, and in the process reduce age quickly. Consecutive occurrences ($\text{flag} == 1$) of this case (tracked by increasing $\gamma$ by $1$ in line~\ref{alg:oneincr}) attempt to decrease backlog even more aggressively, by a larger power of $2$.

The condition that $b_k$ and $\delta_k$ are both negative (line~\ref{alg:four}) corresponds to a reduction in both age and backlog. ACP+ greedily aims at reducing backlog further in the hope that age will reduce too. It attempts MDEC (line~\ref{alg:fourMDEC}) if previously the condition $b_k,\delta_k>0$ was satisfied; otherwise, it attempts the additive decrease DEC.

The source ACP+ targets an increase in average backlog over the next control epoch in case either $b_k>0, \delta_k<0$ or $b_k<0, \delta_k>0$. On the occurrence of the first condition (line~\ref{alg:three}) ACP+ greedily attempts to increase backlog. The condition $b_k<0, \delta_k>0$ hints at too low an update rate causing an increase in age. So, ACP+ attempts an additive increase (line~\ref{alg:two2}) of backlog.


\section{Updates Over Intercontinental Paths and a Contended WiFi Access}
\label{sec:orbit_results}

\begin{figure}[!t]             
\begin{center}
\includegraphics[width=0.48\textwidth]{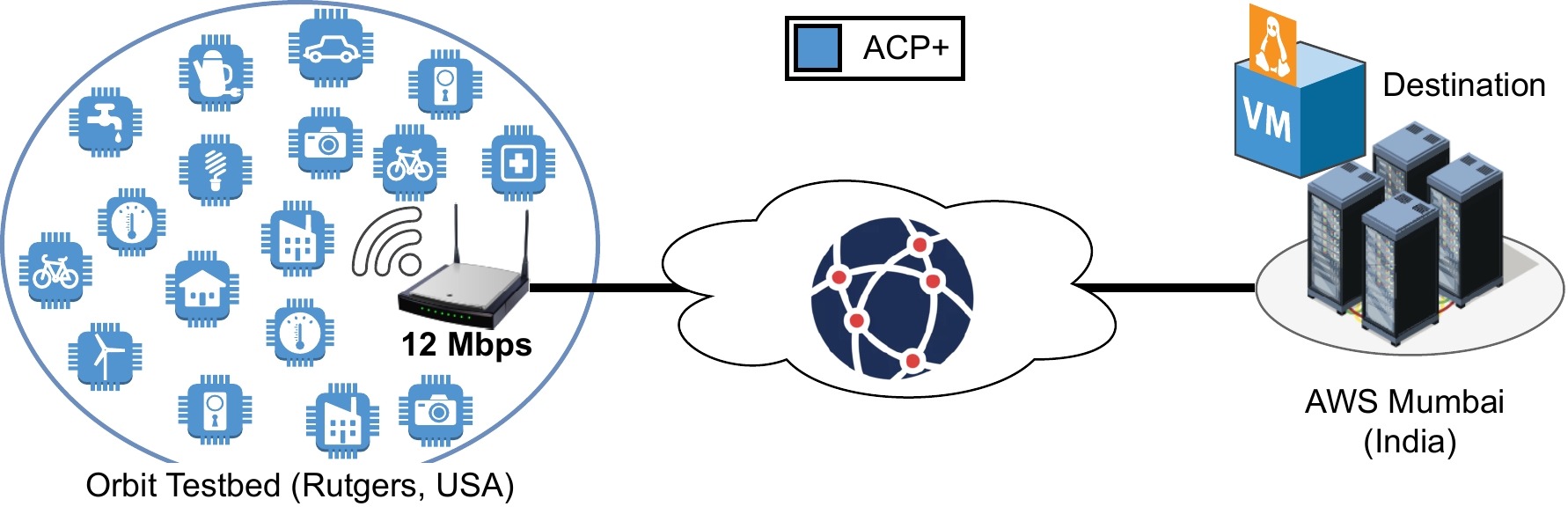}
\caption{ACP+ clients connected to a WiFi AP located in the Orbit Testbed's WiFi grid in USA. They sent updates to a server in AWS Mumbai, India.}
\label{fig:orbit_setup}
\end{center}
\end{figure}

\Cref{fig:orbit_setup} illustrates our real-world experimental setup. We used the ORBIT testbed\footnote{https://www.orbit-lab.org/}, an open wireless network emulator grid located in Rutgers University, USA. The testbed houses multiple wireless capable and programmable radio nodes deployed in a grid fashion with a $1$m separation between adjacent nodes, along columns and rows of the grid. We configured multiple ORBIT nodes as updating sources and an additional ORBIT node as an $802.11$n access point. This access point acts as a gateway to the Internet for our sources and is configured to operate at $5$ GHz on a fixed channel and a fixed WiFi physical layer rate using \texttt{hostapd} and the \texttt{iwconfig} utility. Fixed WiFi rates, in contrast to allowing WiFi rate control, enable better understanding of the impact of the WiFi access and the Internet beyond on the age of updates of the sources at the monitor. Our sources send updates to an ec2 AWS\footnote{https://aws.amazon.com/} instance in Mumbai, India, which serves as our ACP+ monitor. 

For our experiments, we selected up to $80$ nodes (sources) in the testbed to connect to the WiFi access point as its clients. We also configured a node as a \textit{sniffer} to capture packets sent over the WiFi channel. This enabled us to quantify the packet retry rates at the WiFi medium access control layer due to packet collisions or drops over the WiFi access, using the retry flag in the MAC header of sniffed packets. In the end-to-end path, we configured only the wireless network within the ORBIT testbed; the  rest of the path to AWS Mumbai traversed the public shared Internet.
%

%

We experimented with $1$, $2$, $5$, $10$, $20$, $40$, $80$ sources and WiFi physical layer rates of $6$, $12$, and $24$ Mbps. For all our experiments, we estimated the bottleneck link rate over the end-to-end path to be the WiFi link rate. Specifically, in the absence of the WiFi access, the end-to-end path to AWS Mumbai was able to support TCP throughputs as high as $200$ Mbps. Further, the \emph{baseline} RTT between our sources and the monitor is within the $200$-$210$ ms range. It is the average RTT observed by any source when it alone sends to the monitor in a stop-and-wait fashion, i.e., the source sends an update packet and waits for an ACK (or a timeout) before sending the next update packet. The baseline RTT is calculated in the absence of any wireless contention. About $40\%$ of it is speed-of-light propagation delays and the rest is time spent waiting in router queues and transmission times over the hops in the round trip path.

We perform at least five repeats of each experiment configuration, which includes a choice of number of sources, their locations on the ORBIT grid and the WiFi rate. Averages over the repeats are used to evaluate the performance of ACP+. Our experiments lasted over several months and were repeated on different days of the week and at different times of the day.

\begin{figure}[!t]             
	\begin{center}
		\subfloat[Average Age (ms)]	{\includegraphics[width=.245\textwidth]{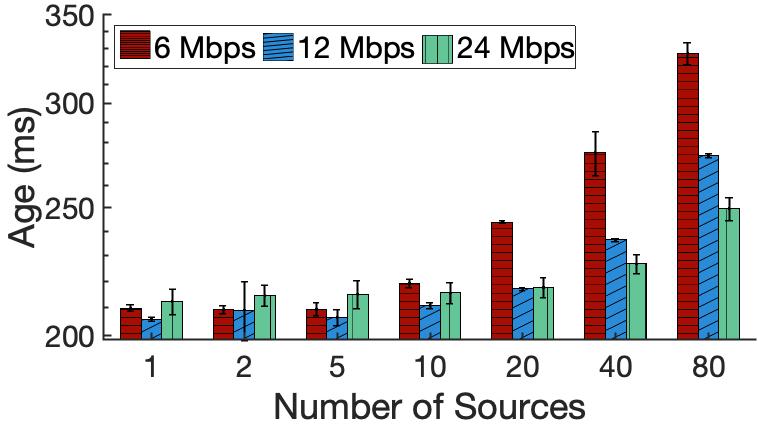}
			\label{fig:age_highContention}}
		\subfloat[Average Backlog (ms)]{\includegraphics[width=.245\textwidth]{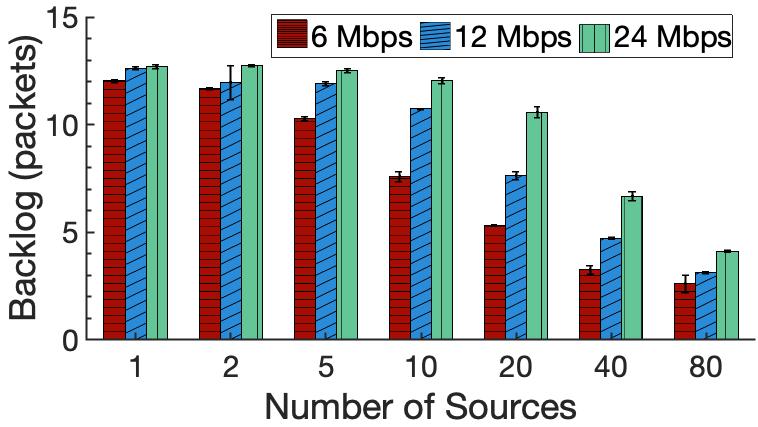}
			\label{fig:bkl_highContention}}
			\linebreak
		\hspace{.001in}
		\subfloat[Throughput (Mbps)]{\includegraphics[width=.245\textwidth]{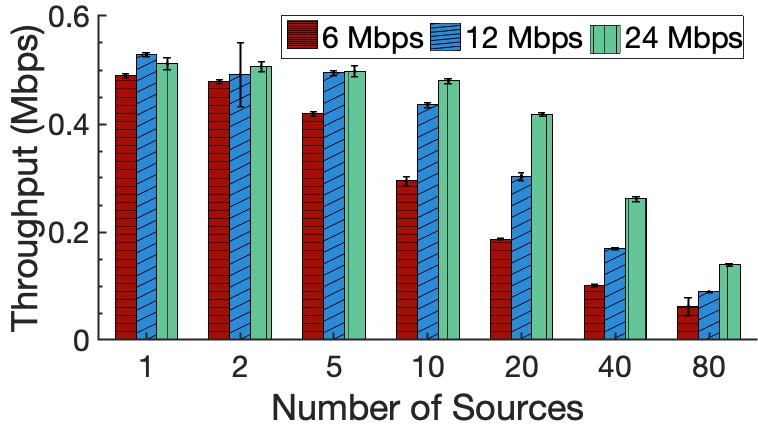}
			\label{fig:thr_highContention}}
		\subfloat[Average RTT (ms)]{\includegraphics[width=.245\textwidth]{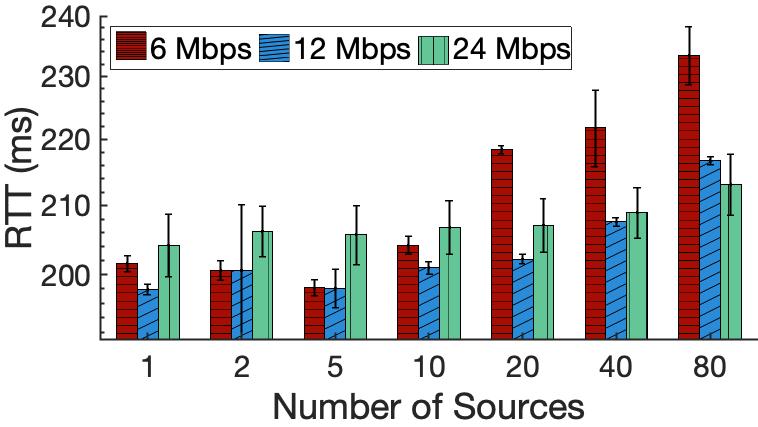}
			\label{fig:rtt_highContention}}	
		\caption{Averages of per source time-average age, backlog, throughput, and RTT, measured over runs of ACP+ for choices of WiFi link rates and number of nodes sharing the WiFi access in the ORBIT testbed.}
		\label{fig:highContentionACP}
	\end{center}
\end{figure}

\subsection{Takeaways for Age Control in the Internet}
For an update payload size of $1024$ bytes, we observe in Figures~\ref{fig:age_highContention} and~\ref{fig:thr_highContention}  that ACP+ achieves a small average age for a single source at an end-to-end throughput of about $0.5$ Mbps, which is much smaller than our chosen WiFi rates. Note that the WiFi link rate is the bottleneck link rate for our paths between the sources and the monitor. The small age-optimizing throughput has been observed by us over other Internet paths (see Section~\ref{sec:ageing_main} and earlier work~\cite{tanya-kaul-yates-wowmom2019}). 

An age optimizing throughput much smaller than the access link rates enables multiple ACP+ sources to share the access without suffering an age penalty because of the other sources. Specifically, Figure~\ref{fig:age_highContention} shows that the average age per source stays similar when there are $1-5$ sources sharing a $6$ Mbps WiFi access. The age also stays similar for  $10$ and $20$ sources when sharing, respectively, a $12$ and $24$ Mbps WiFi access. As the number of sources increase beyond $5$, $10$, and $20$ sources, respectively for access link rates of $6$, $12$, and $24$ Mbps, the increased contention results in a rapid increase in age with the number of sources.

The increased contention results in large RTT(s) and this prompts ACP+ to maintain smaller backlogs of updates per source. The RTT(s) and corresponding backlogs are shown in Figures~\ref{fig:rtt_highContention} and~\ref{fig:bkl_highContention}, respectively. The rapid increase in RTT per source with increasing contention, and the smaller backlogs, result in a sharp reduction in per source throughputs in Figure~\ref{fig:thr_highContention}.

Age-optimizing throughputs much smaller than access link rates have significant consequences for age control over the Internet, as we demonstrated using ACP+. It implies that multiple sources can share the access, which is the bottleneck link in the path, without much contention and without saturating the shared access. This behavior is contrary to that of TCP, which always saturates the bottleneck link, irrespective of the number of senders sharing it. 

For a small enough number of sources sharing an access, age optimization is constrained by the backhaul beyond the access. While the backhaul has a bottleneck link rate much larger than that of the WiFi access, the time that an update spends buffered in the many hops that constitute the backhaul, given the other traffic using it, is the most salient as regards age control. %
\RY{On the path to AWS Mumbai, do we have any idea how much of the delay is (1) buffering, (2) total transmission time over multiple hops, (3) speed of light propagation? My back of the envelop calc based on a fiber path from NJ to Mumbai being 7500 miles is that one way propagation time in fiber is 59ms. So, out of a 200ms RTT, about half is propagation?}\RSP{Added the percentages to the third para of this section.} %
As the number of sources sharing an access gets large, saturation of the access link becomes the constraining factor with regards to age optimization.

\begin{figure}[!t]             
\begin{center}
\includegraphics[width=0.75\linewidth]{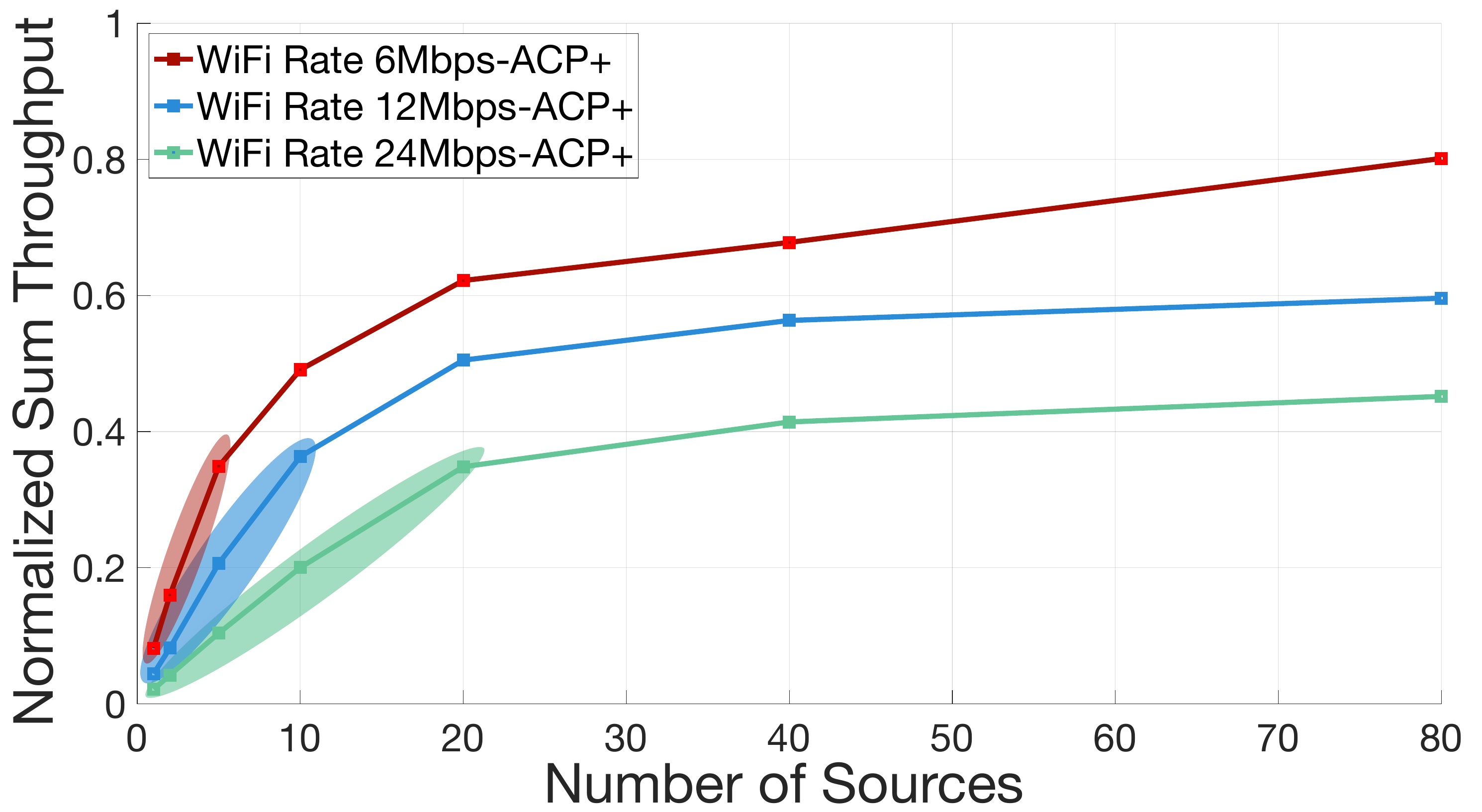}
\caption{Sum Throughput of ACP+ normalized with respect to the bottleneck link. The shaded ellipse shows the region where the access is not the bottleneck for the timeliness performance of ACP+. Beyond the shaded region, the access becomes the bottleneck, and the sum throughputs start saturating.}
\label{fig:SumThroughput_barACP}
\end{center}
\end{figure}

Figure \ref{fig:SumThroughput_barACP} illustrates the two regimes of age control. It plots the utilization of the access channel (measured as the sum of throughputs of the sources sharing it) normalized by the WiFi link rate. When the number of sources sharing the access is small and contention between the sources is low, the normalized utilization of the access increases in proportion (marked by the shaded ellipses) to the number of sources. The access isn't the constraining factor and age control must adapt to the utilization of the backhaul by other traffic, while ensuring that a large enough backlog of source updates is maintained in the backhaul, given the large number of hops that may constitute it. Beyond the region of low contention, the normalized utilization flattens and converges close to the maximum obtainable for the link rate\footnote{The normalized utilization is less than $1$ because of WiFi protocol overheads including headers of layers $1$ and $2$. The overheads are larger for larger link rates.}. Not surprisingly, the region of low contention extends to a larger number of sources for  larger WiFi link rates.

We could not extend our experiments to include greater than $80$ nodes on the ORBIT testbed due to hardware and wireless driver restrictions. As a result we couldn't explore the region of really high contention, given age optimizing throughputs, wherein one would expect age control to ideally backlog on an average less than an update per source per round-trip-time\footnote{The older version ACP~\cite{tanya-kaul-yates-wowmom2019} maintained backlogs of about $1$ per source as contention increased.}. %
\RY{What about larger update payloads to increase offered load/contention?}\RSP{Added a footnote.} 
We will resort to simulations to evaluate how ACP+ performs when the contention resulting from the sources sharing the access is very high\footnote{Note that increasing the update size to $1500$ bytes wouldn't increase contention, given that nodes in the ORBIT grid are within sensing range of each other.}. 

\section{Simulations Setup and Results}
\label{sec:evaluation_acpNew}
\begin{figure}[!t]             
	\begin{center}
		\includegraphics[width=0.49\textwidth]{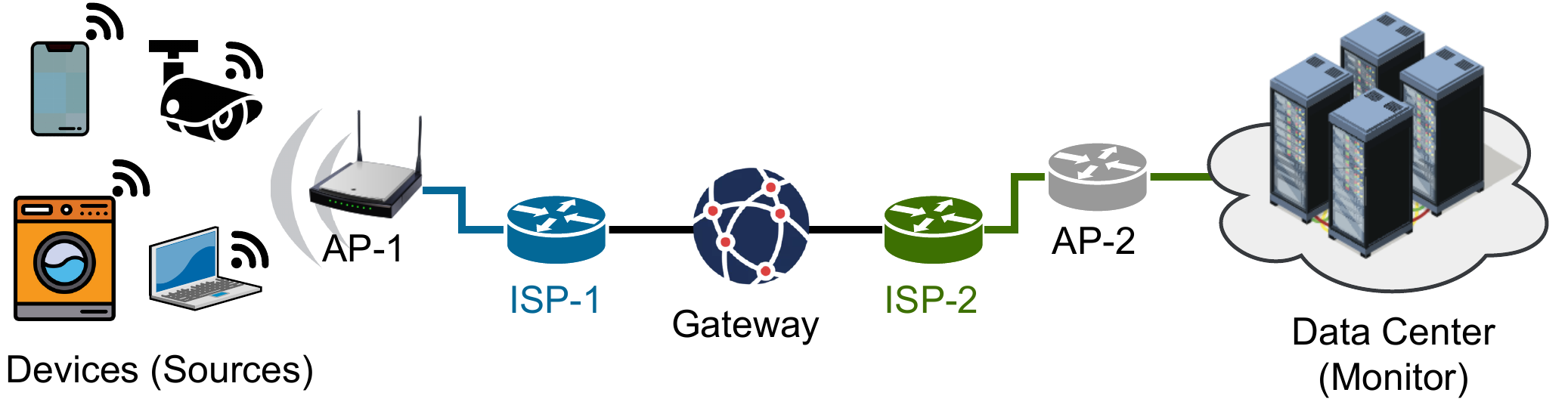}
		\caption {Sources are connected to the monitor via multiple routers and access points. Each source update travels over six hops. The first hop is between the source and access point AP-1. This could be either P2P or WiFi. The other hops that involve the ISP(s) and the Gateway are an abstraction of the Internet. These hops are P2P links and we vary their rates to simulate different end-to-end RTT.}
		\label{fig:simulationNetwork}
	\end{center}
\end{figure}

Figure~\ref{fig:simulationNetwork} shows the end-to-end network used for simulations. 
We performed experiments for $1-48$ sources accessing AP-1 using the WiFi (802.11g) medium access. We simulated for sources spread uniformly and randomly over an area of $20\times 20$ m$^2$. The channel between a source and AP-1 was chosen to be log-normally distributed with choices of $4$, $8$, and $12$ for the standard deviation. The pathloss exponent was $3$. 
We used the network simulator ns3\footnote{\url{https://www.nsnam.org/}} together with the YansWiFiPhyHelper\footnote{\url{https://www.nsnam.org/doxygen/classns3_1_1_yans_wifi_phy.html}}. Our simulated network is however limited in the number of hops, which is six.
WiFi physical (PHY) layer rate was set to $12$ Mbps and that of the P2P links was set to $6$ Mbps. 

To compare the age control performance of ACP+, we use \emph{Lazy} as defined in our earlier work~\cite{tanya-kaul-yates-wowmom2019}. \emph{Lazy}, like ACP+, also adapts the update rate to network conditions. However, it is very conservative and keeps the average number of update packets in transit small. Specifically, it updates the $\overline{\text{RTT}}$ every time an ACK is received and sets the current update rate to the inverse of $\overline{\text{RTT}}$. Thus, it aims at maintaining an average backlog of $1$.


\begin{figure}[!t]             
	\begin{center}
		\subfloat[]{\includegraphics[width=.45\linewidth]{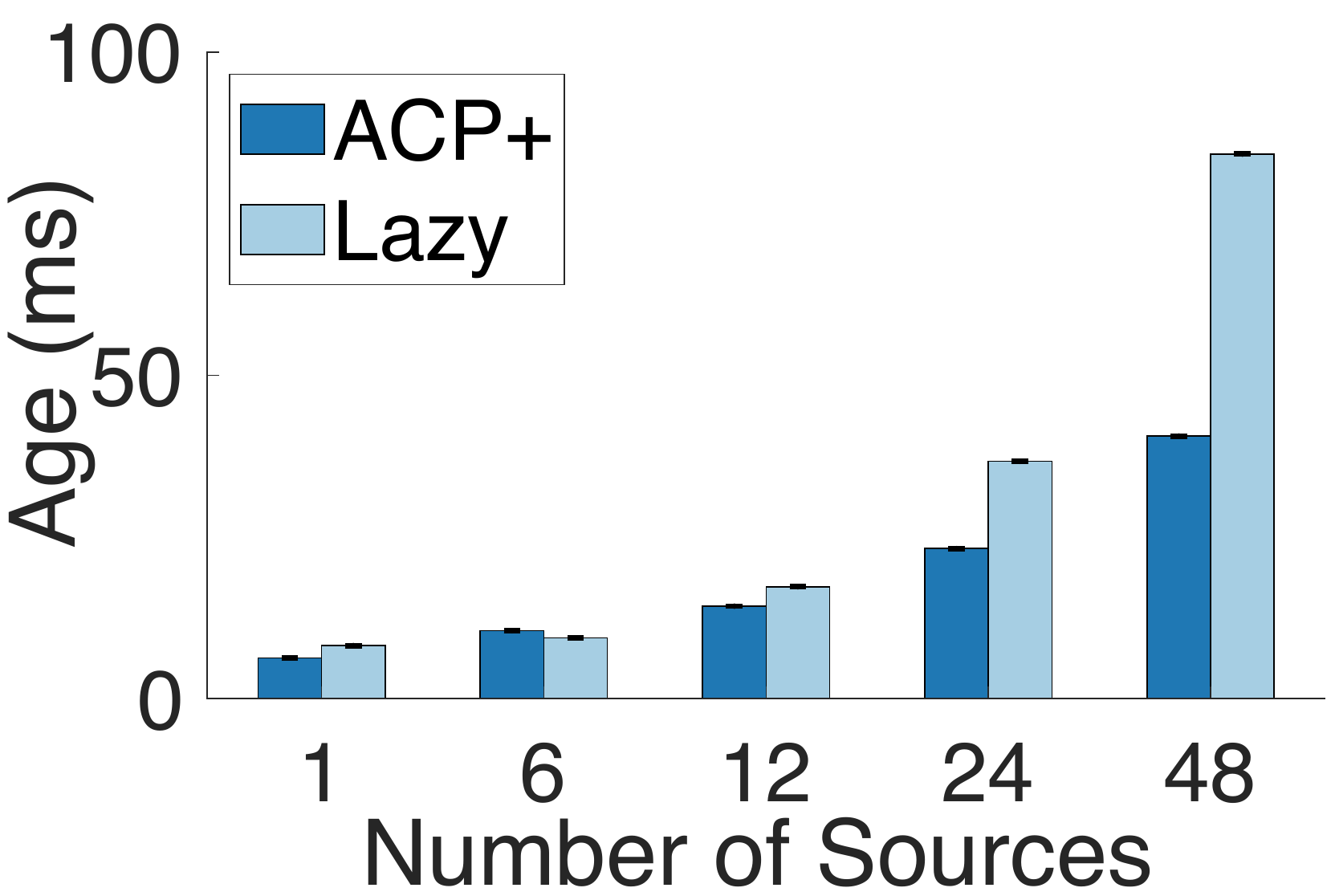}
			\label{fig:age_topology_1}}
		\subfloat[]{\includegraphics[width=.5\linewidth]{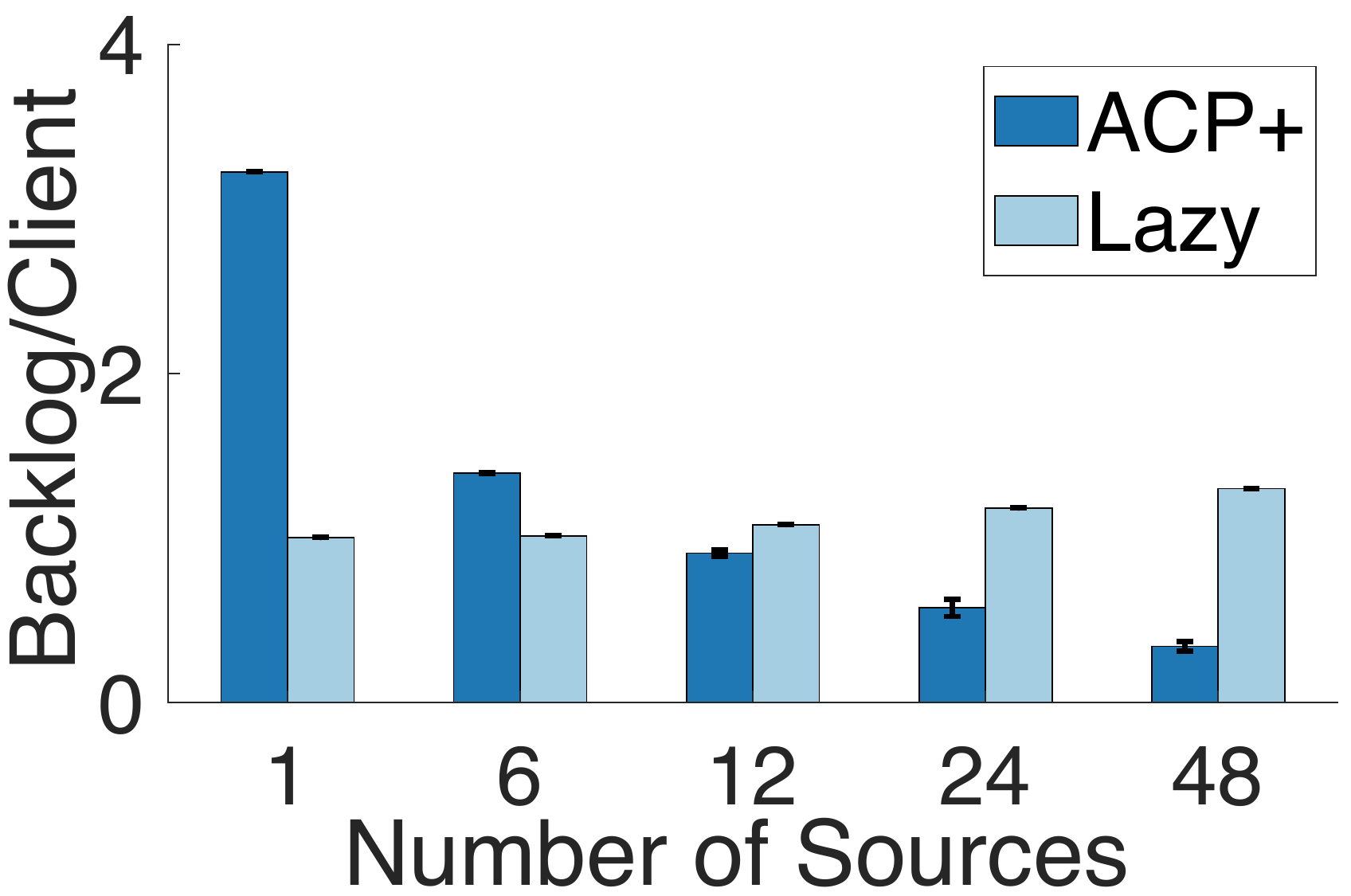}
			\label{fig:backlog_topology_1}}
            \caption{(a) Average source age and (b) Average source backlog for \emph{Lazy} and ACP+ when all links other than wireless access are $6$ Mbps. All sources used a WiFi PHY rate of $12$ Mbps. The sources are spread over an area of $400$ m$^2$.}
		\label{fig:ageAndBacklogFastAndSlowNets_new}
	\end{center}
\end{figure}

Figure~\ref{fig:age_topology_1} shows that ACP+ achieves a smaller age per source than \emph{Lazy}. The improvements are especially significant when a large number of sources share the access to AP-1. That ACP+ is able to achieve smaller ages can be understood via the average backlog per source when using ACP+ and \emph{Lazy}, which is shown in Figure~\ref{fig:backlog_topology_1}. When we have just one source, ACP+ tries to fill each queue in the network with an update. Note that, unlike the Internet, this network isn't shared by any other traffic. This results in a larger backlog and a lower age in comparison to \emph{Lazy}, which achieves a backlog of just $1$ update. However, as the number of sources increases, while \emph{Lazy} continues to maintain a backlog of $1$ per source, ACP+ reduces it. The backlogs obtained are $3.23, 1.39, 0.91, 0.57, 0.34$, respectively, for $1,6,12,24,48$ sources. 

The ACP+ backlogs when we have a large number of sources are not only much smaller than \emph{Lazy}, it turns out that they are not too far from an ideal scheduling mechanism that schedules updates from the sources in a round-robin and contention free manner. Note that in the absence of contention and packet drops due to channel errors and given deterministic transmission times that are the same for updates from all source, round-robin optimizes the average age~\cite[Lemma 4]{kadota-SUSM-ton2018}. %
To see this, consider the simplified setting in which the WiFi and P2P links are the same and no packets are dropped due to channel errors over WiFi. A round-robin scheduler would keep six updates in transit of the source when we have just one source. 
This would result in a backlog of $6$. It would schedule six sources one after the other in a manner such that a round of scheduling would lead to six packets in the six queues from the six different sources, resulting in an average backlog of $1$ per source. Similarly, for when we have $12, 24, 48$ sources, we would see backlogs per source of $1/2, 1/4, 1/8$, respectively. ACP+ sees larger backlogs than these, at least partly because of packet collisions over WiFi access, which results in larger delays in WiFi hop.

\emph{Age Fairness Using ACP+}: 
We quantify fairness in age achieved by multiple ACP+ sources that share an access network and send their updates to a monitor. 
We use the Jain's fairness index~\cite{jain1984quantitative} to quantify \textit{age fairness} in both our simulations and real-world experiments. The Jain's fairness index can take values between $0$ and $1$. A larger fairness index implies more similar ages of the different sources at the monitor. An index of $1.0$ indicates that the ACP+ control algorithm enables the ACP+ flows to achieve the same ages over their paths to the monitor.

In our simulations, we find that as we increase the number of sources sharing the network from $6$ to $48$, our fairness index reduces from $.99$ to $.89$. In our real-world experiments, the fairness index lies between $.99$ to $1.0$ as we increase source density from $2$ to $80$ for all the WiFi link rates ($6$, $12$ and $24$ Mbps). ACP+ ensures age fairness in our experiments.

\section{Congestion Control Approaches in the Internet: Do we need ACP?}
\label{sec:ageing}
TCP is still the dominant protocol used in the Internet with $\approx 90\%$ traffic share~\cite{ruth2018first}. TCP congestion control is the primary mechanism by which end hosts share available Internet bandwidth. 

\begin{figure}[t]             
	\begin{center}
	        \subfloat[RTT as a function of offered load]{\includegraphics[width=.47\textwidth]{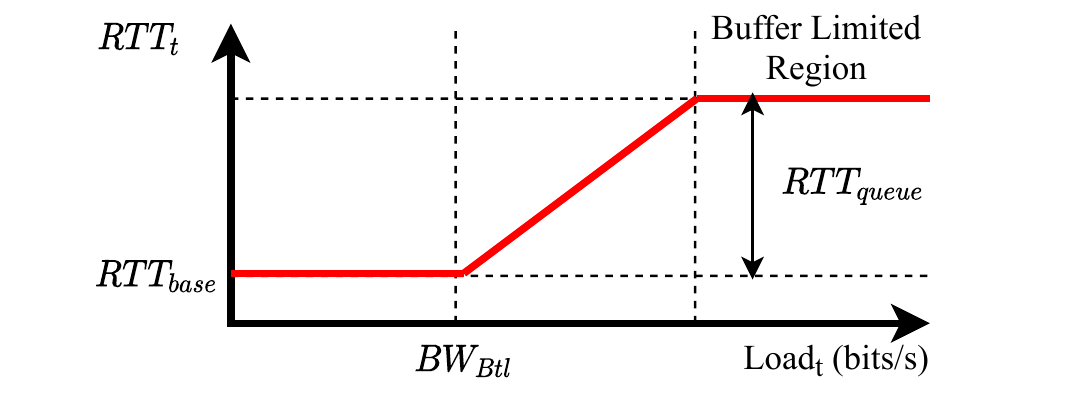}
			\label{fig:transient_tcp}}
			\\
			\subfloat[ Average steady state RTT as a function of average load]{\includegraphics[width=.47\textwidth]{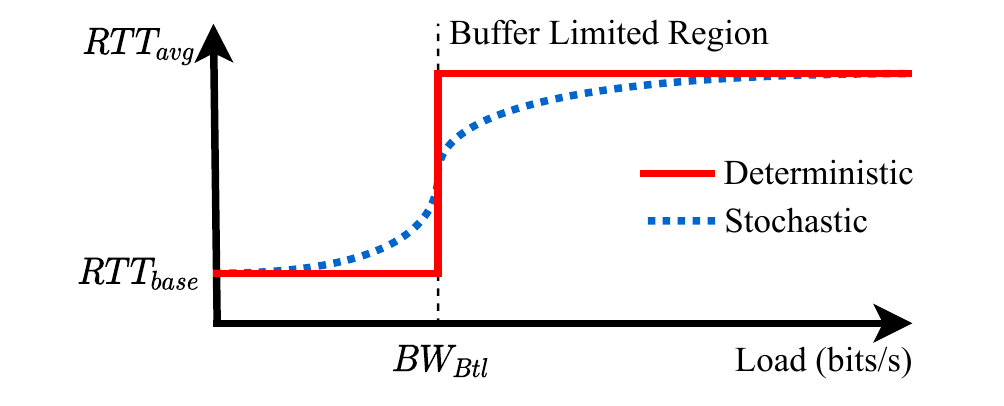}
			\label{fig:steady_tcp}}
		\caption{An illustration of how round-trip times vary as a function of the offered load. While (a) shows the change in instantaneous RTT as the load increases, (b) shows  the steady-state average behavior at a chosen load.}
		\label{fig:TCPtransient_SteadyState}
	\end{center}

\end{figure}
For the purpose of TCP's operation, the end-to-end path  may be abstracted away as a link with bottleneck bandwidth $BW_{\text{Btl}}$ and a round-trip propagation time of $RTT_{\text{base}}$ (baseline RTT)~\cite{cardwell-bbr-2016}. Figure~\ref{fig:transient_tcp} provides an illustration, akin to that in~\cite[Figure $1$]{cardwell-bbr-2016}, of the \emph{instantaneous} round-trip time $RTT_t$ at time $t$
as a function of the current offered load (the effective rate at which TCP is sending bytes).  As long as the offered load is smaller than $BW_{\text{Btl}}$, the TCP packets see a low and fixed RTT of $RTT_{\text{base}}$. Once the offered load becomes larger than $BW_{\text{Btl}}$, the TCP packets that arrive at the link's queue see increasingly more packets waiting for service ahead of them. This results in a linear increase in $RTT_t$ until the queue becomes buffer limited, the RTT saturates and TCP packets arriving at a full queue are dropped. 

Traditionally, TCP's loss-based congestion control allows for an increasing number of unacknowledged bytes from an application to flow through the network pipe until one or more bytes sent are lost. Loss of bytes signals congestion to the TCP sender and occurs because a link along the end-to-end path is operating in the buffer limited region. An end-to-end flow may achieve a throughput equal to the bottleneck bandwidth, but packets in the flow will suffer large round-trip times, especially when the link has a large buffer. 

Figure~\ref{fig:transient_tcp} suggests that one would ideally like to operate at the lower ``knee'' in the curve, i.e., close to the bottleneck throughput $BW_{\text{Btl}}$ but at low delays. In fact, delay-based and hybrid congestion control algorithms such as the recently proposed \emph{Bottleneck Bandwidth and Round-trip propagation time} (BBR)~\cite{cardwell-bbr-2016} protocol, attempt this by using the round-trip time to detect congestion early, before a loss occurs due to buffer unavailability at a certain router along the path. Intriguingly, this point of operation would satisfy the goal of age control by resulting in the highest rate of packet delivery at the monitor at the smallest possible packet delays. In fact this combination of a throughput of $BW_{\text{Btl}}$ and round-trip times of $RTT_{\text{base}}$ is achieved by the snapshot in Figure~\ref{fig:goodACP_03} that illustrates a good age control algorithm in action. 



Of course, as was observed in~\cite{kleinrock2018internet} in relation to the stated point of operation of BBR, when a path is better modeled by a stochastic service facility, the average round-trip times at the maximum achievable throughput of $BW_{\text{Btl}}$ could be much larger than $RTT_{\text{base}}$. Figure~\ref{fig:steady_tcp} provides an illustration of \emph{steady-state} average RTT as a function of average load. The red and blue curves, respectively, correspond to a deterministic and a stochastic service facility.

\begin{figure}[!t]             
\begin{center}
\includegraphics[width=0.9\linewidth]{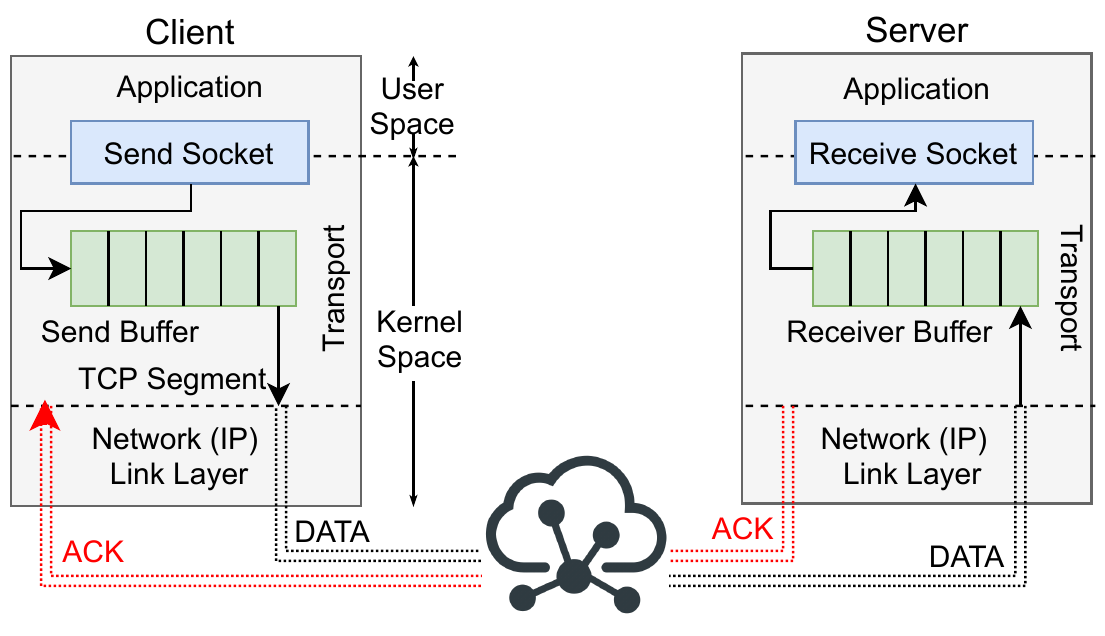}
\caption{ TCP Network Stack}
\label{fig:TCP-Stack}
\end{center}
\end{figure}

The shift in congestion control algorithms from keeping the pipe full to ``keeping the pipe just full, but no fuller''~\cite{kleinrock2018internet}, motivates this empirical study of how the information at a receiver would age if updates were transmitted over the cloud using the congestion control algorithms. We evaluate a mix of loss-based (Reno~\cite{reno} and CUBIC~\cite{ha2008cubic}), delay-based (Vegas~\cite{brakmo1994tcp}) and hybrid congestion control algorithms (YeAH~\cite{baiocchi2007yeah} and BBR~\cite{cardwell-bbr-2016}). 

However, we must be careful as (a) TCP doesn't regulate the rate of generation of packets by the status updating application, (b) it is a stream-based protocol and has no notion of update packets. As illustrated in Figure~\ref{fig:TCP-Stack}, an application  writes a stream of bytes to the TCP sender's buffer. TCP creates segments from these bytes in a first-come-first-serve manner. TCP segments are delivered to the TCP receiver. At any time, TCP allows a total of up to a current congestion window size of bytes to be in transit in the network. The TCP receiver sends an \texttt{ACK} to inform the sender of the last segment received.

To stay focused on evaluating how 
scheduling TCP segments over an end-to-end path would age updates at a receiver, we assume that a TCP segment, when created, contains fresh information. Specifically, we ignore the ageing of bytes while they wait in the TCP send buffer. One way of achieving this in practice would be to have the application provide freshly generated information (a generate-at-will model~\cite{yates2020age-survey}) to be incorporated in a segment just as TCP schedules it for sending.

We approximate the age of the segment when it arrives at the TCP receiver to be the RTT of the segment, which is calculated based on the time of receipt of the TCP \texttt{ACK} that acknowledges receipt of the segment. Further, we approximate the inter-delivery time of segments at the receiver by the inter-delivery times of the corresponding \texttt{ACKs}. The RTT(s) and the inter-delivery times together allow us to come up with an estimate of the time-average of the age function (see Figure~\ref{fig:ageSampleFunction}) at the receiver that results from a congestion control algorithm. 

Last but not the least, we would like to minimize the impact of packet loss due to link transmission errors on our evaluation of congestion control. Given our focus on paths in the cloud, specifically between AWS data centers, we observe a very small percentage of loss, and that too because of buffer overflows in routers that result in the process of congestion control estimating the bottleneck bandwidth.
\section{Ageing over the Internet}\label{sec:ageing_main}

We empirically determine the ability of TCP congestion control algorithms to deliver \textit{fresh} updates over an end-to-end Internet path. We also compare its performance with ACP+. First, we focus on the core network. The Internet core is widely regarded to be significantly reliable (as also seen in our experiments described later) and is operated by managed entities such as the Amazon Web Services (AWS). Next, we look at an end-to-end path in which the first hop is a shared and contended wireless access. 

\subsection{Ageing over the Core Network}\label{sec:awsCore}
\begin{figure}[!t]             
\begin{center}
\includegraphics[width=0.95\linewidth]{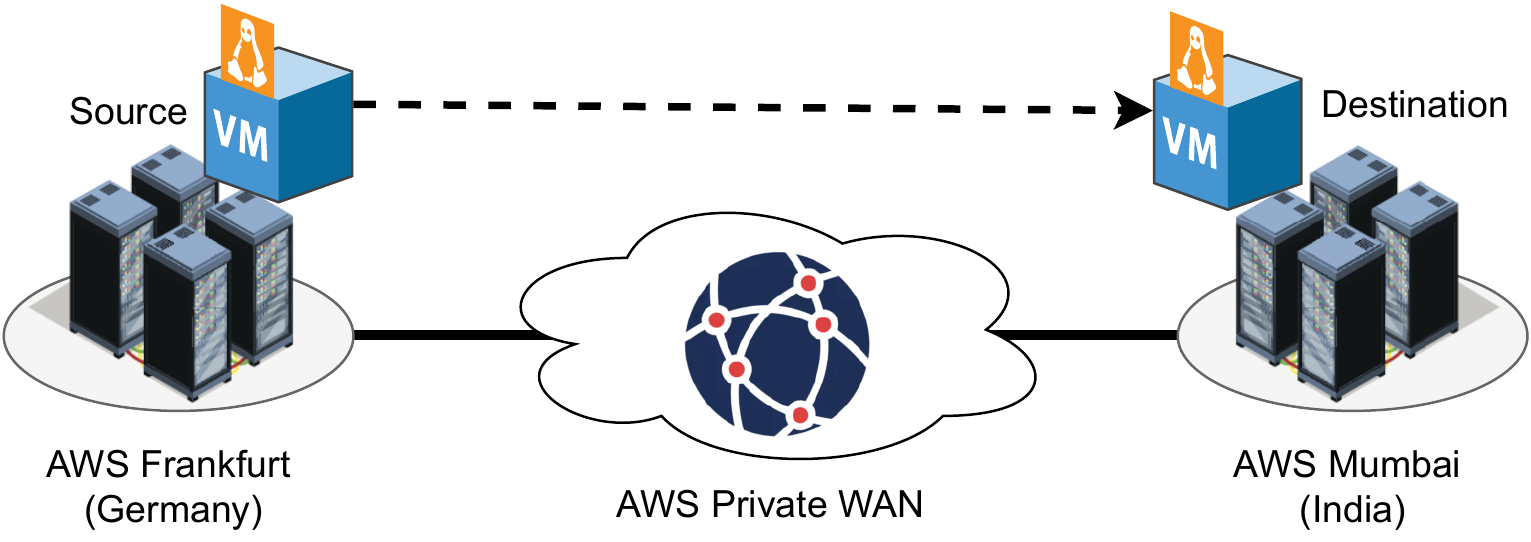}
\caption{An illustration of the real experiment topology on the AWS EC2 cloud network. The client machine (both ACP+/TCP) was in AWS Frankfurt, Germany, and the server was in AWS Mumbai, India. The instances were connected via the AWS Private WAN.}
\label{fig:AWS-setup}
\end{center}
\end{figure}

Figure~\ref{fig:AWS-setup} shows the real experiment topology. All our experiments over the Internet used two T2.micro instances in the AWS EC2 cloud network. Both instances are configured with one virtual CPU, $1$ GB RAM and a $1$ Gbps Ethernet link connected to the AWS private WAN. One of the instances was in the AWS Frankfurt (Germany) data-center, while the other was deployed in the AWS Mumbai (India) data-center. Each instance ran a virtual machine with Ubuntu $18.04$ LTS with Linux kernel version $5.3$. We confirmed through periodic \texttt{traceroute} that the underlying network between our two chosen instances was served by the AWS private WAN. 

For both the ACP+ and TCP experiments, we deployed the sender in AWS Frankfurt and the receiver in AWS Mumbai. 
For each chosen congestion control algorithm, we investigated the impact of different receive buffer sizes on the performance of the congestion control algorithms by changing \texttt{default} and \texttt{maximum} values of \texttt{r\_mem} in the Linux kernel. The space available in the receiver buffer limits the maximum amount of bytes that any congestion control algorithm may send to the TCP receiver.

For the TCP experiments, we used \texttt{iPerf3} for packet generation and \texttt{Wireshark} for packet captures. To ensure that all algorithms saw similar network conditions we ran multiple iterations of ACP+, TCP BBR, TCP CUBIC, TCP Reno, TCP Vegas and TCP YeAH, in that exact order, one after the other. For each TCP variant in the stated order, we further ran different receive buffer settings. Each run of the experiment lasted $200$ s.
Considering that end-to-end RTT is $\approx 110$ ms in our setup, 
TCP spends a majority of the transfer time in the steady-state phase. Lastly, we observe that in our experiments the age performance of TCP congestion control algorithms does not take a hit because of TCP's feature of guaranteed in-order delivery\footnote{TCP guarantees delivery of bytes sent by an application. As a result, TCP retransmits lost segments, which might contain stale information. We ignore retransmitted segments for calculation of the time-average age. However, retransmissions and delays incurred because TCP ensures in-order delivery of bytes to the receiving application can result in a large age. Since our goal is to understand the behavior of congestion control algorithms, we would like to minimize the impact of guaranteed in-order delivery on age achieved by the algorithms. Luckily, the core network provides a very reliable byte pipe. Even our measurements over the controlled wireless testbed, which we detail later, saw very few losses ($< 0.1\%$) for up to $80$ nodes. We also measured the duplicate ACKs, which indicate losses and out-of-order packets received at the receiver. We observed very low percentages of duplicate ACKs ($<1\%$) for up to $20$ nodes and up to $\approx 1.5\%$ for up to $80$ nodes. We matched the time of reception of dupACKs to retransmitted data to establish a correspondence. Similar to our duplicate ACKs, the retransmitted data percentages are very low ($\approx 0.5\%$) for up to $80$ nodes.}.


\subsubsection{Results over the Core Network}
We show results from $40$ runs each of ACP+, BBR-d1m1, BBR-d1m3, BBR-d5m5, CUBIC, Reno, Vegas and YeAH. For each run, we show the average age, throughput, and average delay (round-trip time). In the above list, we have BBR run with three different receiver buffer settings. BBR-d1m1 denotes the smallest \texttt{default} and \texttt{maximum} values of the receiver buffer (\texttt{r\_mem}). In BBR-d1m3, the \texttt{default} is the same as BBR-d1m1 but the \texttt{maximum} is three times larger. Similarly, in BBR-d5m5 both the \texttt{default} and the \texttt{maximum} is five times that in BBR-d1m1. For all other TCP algorithms, the results are shown for a \texttt{default} and a \texttt{maximum} five times that of BBR-d1m1. In general, one would expect a larger receiver buffer to allow the TCP algorithm to have a larger number of bytes in flight as long as the network doesn't become the bottleneck.

\begin{figure}[!t]             
\begin{center}
\includegraphics[width=0.85\linewidth]{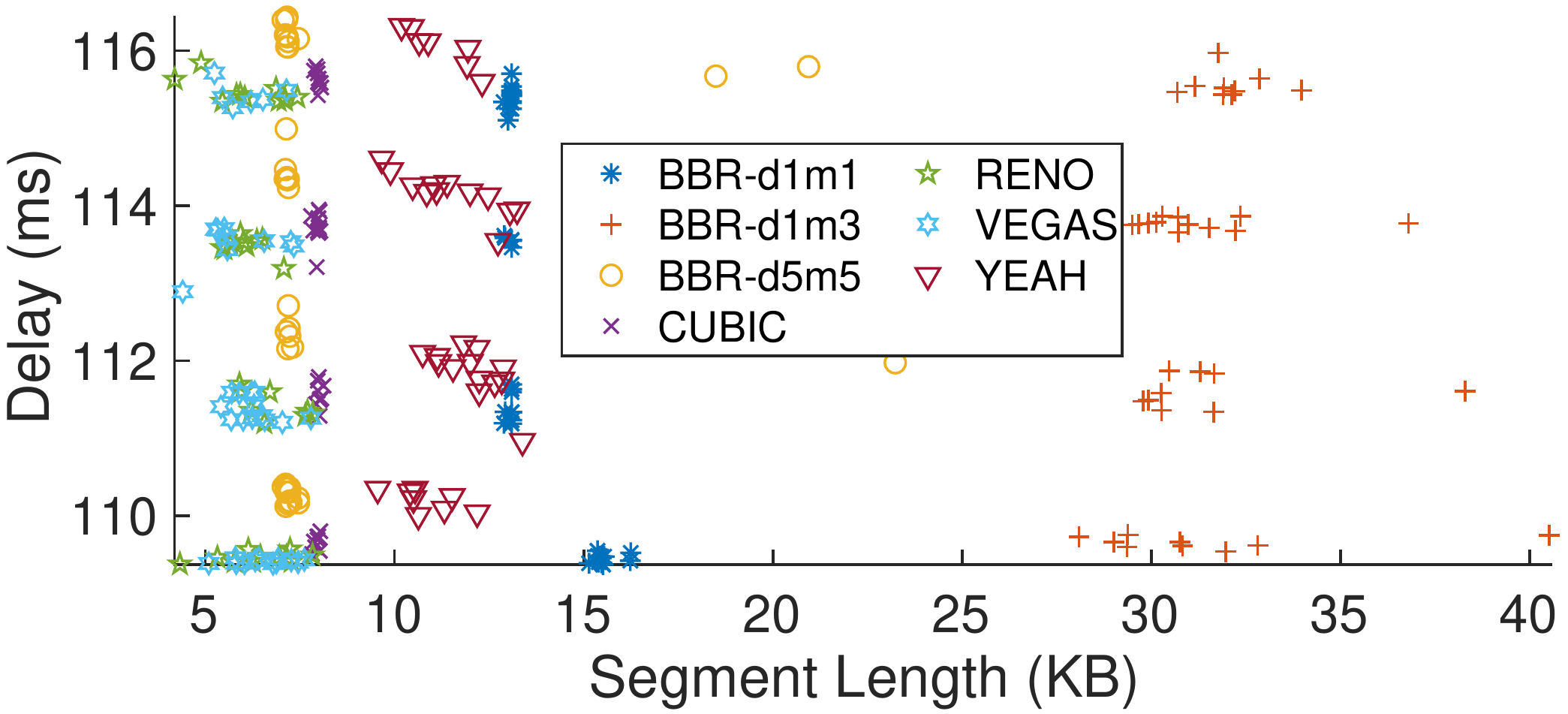}
\caption{TCP segment length vs. delay obtained for the runs of the different algorithms.}
\label{fig:delay_segLen_all}
\end{center}
\end{figure}

\subsubsection{Queue Waiting Delays Dominate} Figure~\ref{fig:delay_segLen_all} shows the impact of TCP segment lengths on delay. As is seen, segment length and delays are uncorrelated 
 for all the TCP algorithms. This observation can be explained by the fact that the delays in the network are almost entirely because of the time spent in router queues awaiting transmission. The transmission times (propagation delays), which are about $20$ ms, are a small fraction in comparison.
It may be worth noting that the TCP segment lengths are chosen by the TCP algorithm and often change during a TCP session. In the figure, we show segment lengths averaged over a run.


\subsubsection{Delay vs. Age} 
\begin{figure}[!t]             
\begin{center}
\includegraphics[width=0.85\linewidth]{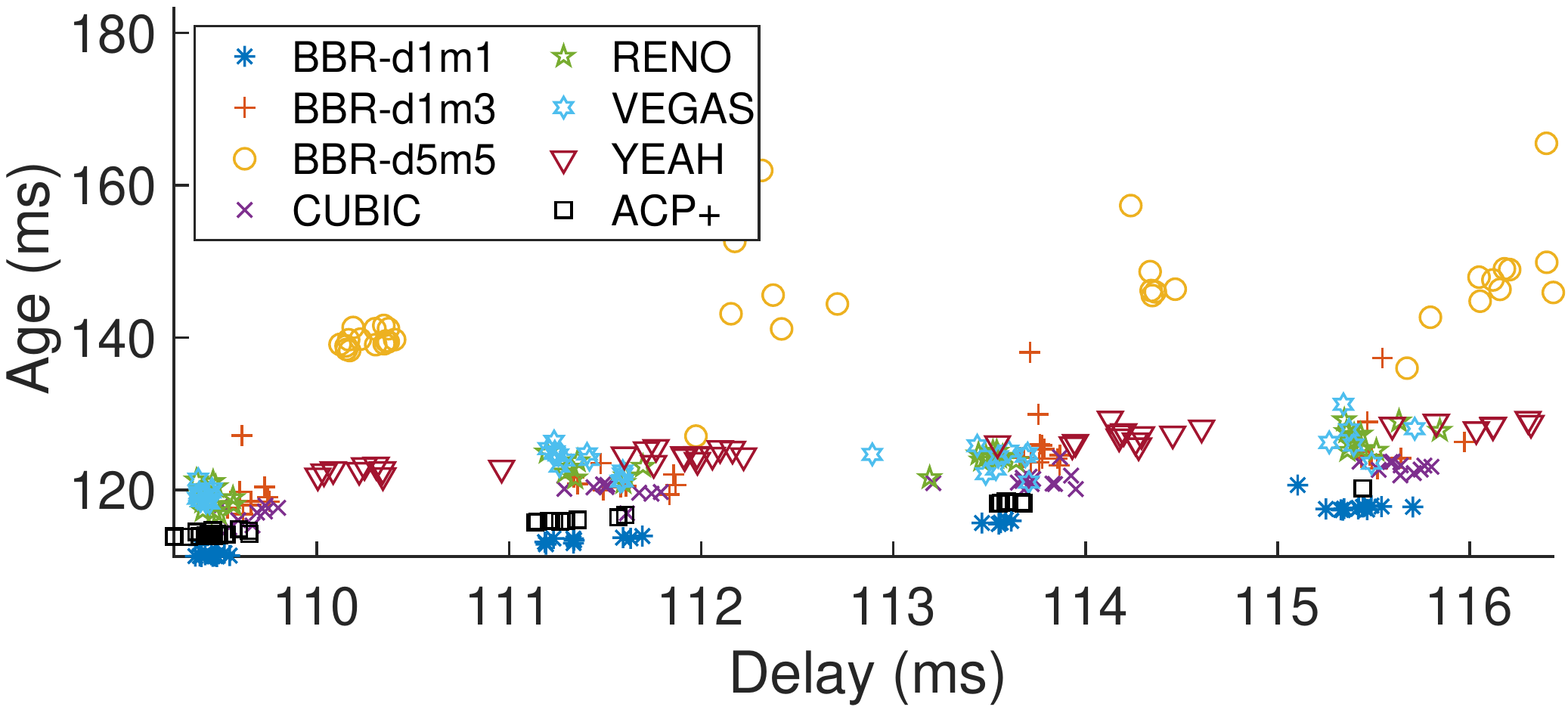}
\caption{Delay vs. age for the different runs of the chosen algorithms.}
\label{fig:delay_age_all}
\end{center}
\end{figure}

Figure~\ref{fig:delay_age_all} shows a scatter of (delay, age) for the chosen runs. We see that BBR-d5m5 sees both age and delays larger than the rest. Amongst the rest, from the figure, it is apparent that ACP+ achieves delays and ages smaller than all algorithms other than BBR-d1m1. BBR-d1m1 achieves a slightly smaller age than ACP+. 

In fact, the age and delay achieved by BBR-d1m1, averaged over all runs, are $114.5$ ms and $112.33$ ms, respectively. The corresponding values for ACP+ are $115.5$ ms and $110.79$ ms. The next smallest age is achieved by CUBIC and is $\approx 121$ ms. Reno, Vegas and BBR-d1m3 achieve higher ages than CUBIC, with YeAH achieving the highest age of about $125$ ms among them. BBR-d1m4, BBR-d1m5 and BBR-d5m5 achieve ages larger than $140$ ms. Only BBR-d5m5 is shown.

\subsubsection{ACP+ vs. BBR-d1m1}
\begin{figure}[!t]             
\begin{center}
\includegraphics[width=0.85\linewidth]{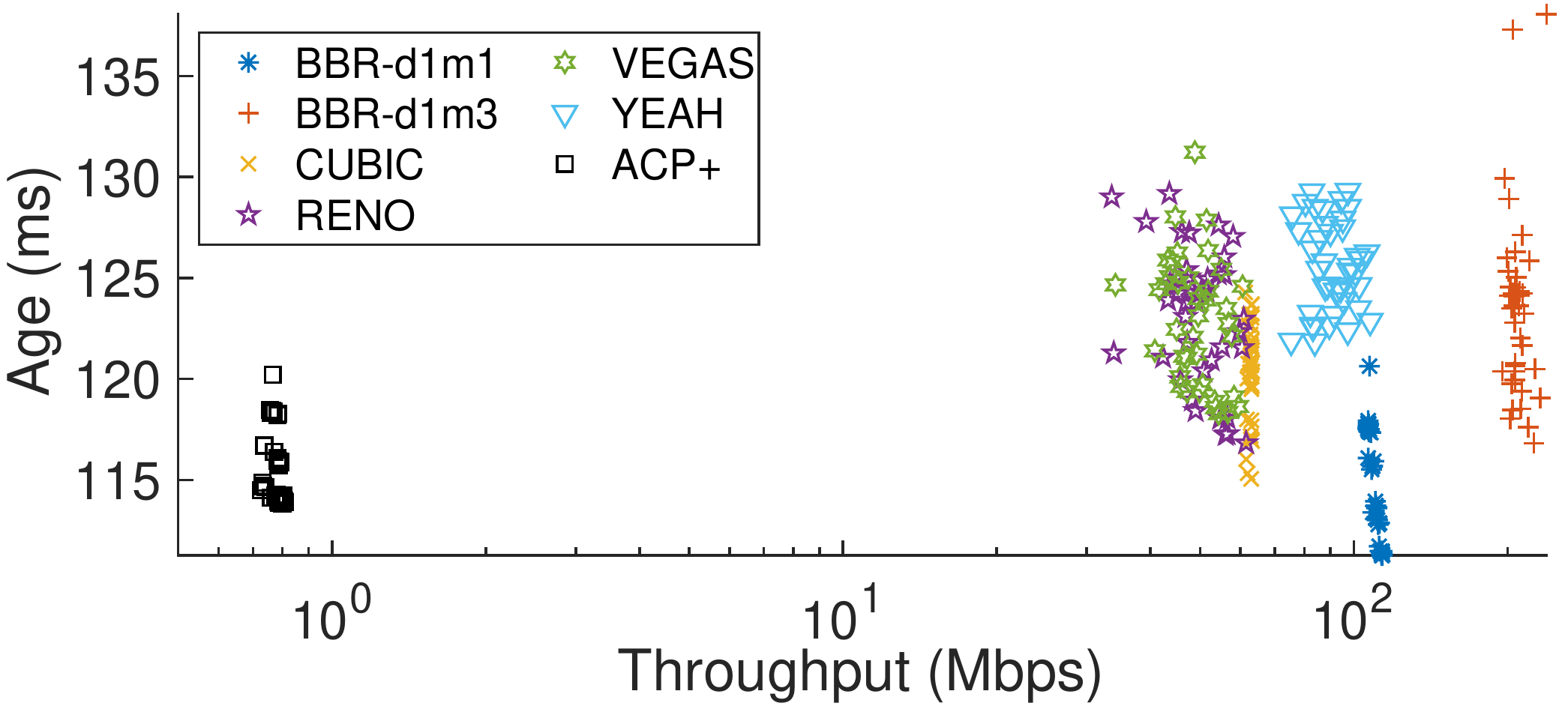}
\caption{Throughput vs. age for the different runs of the chosen algorithms.}
\label{fig:thr_age_all}
\end{center}
\vspace{-0.2in}
\end{figure}

Before we delve further into the relative performances of ACP+ and BBR-d1m1, let's consider Figure~\ref{fig:thr_age_all} in which we show the (throughput, age) values achieved by the different algorithms. We omit BBR-d5m5 from the figure as it resulted in high age values (average larger than $140$ ms) and also did not yield very good throughput. BBR-d1m3 achieves the highest throughput. In fact, its throughput of about $200$ Mbps is twice the next highest value of about $110$ Mbps achieved by BBR-d1m1. The average age when using BBR-d1m3 is $123.5$ ms in contrast to the $114.5$ ms obtained when using BBR-d1m1. 

Interestingly, the throughput obtained by ACP+ is a low of $0.77$ Mbps in contrast to $110$ Mbps obtained using BBR-d1m1 ($\approx 141 \times$ the ACP+ throughput). This stark difference is partly explained by the segment\footnote{Recall our assumption that every new segment contains a fresh update.} sizes used by BBR-d1m1, on an average about $14$ KB, in comparison to the constant $1024$ byte payload of an ACP+ packet. 
This difference still leaves an unexplained factor of about $10$. This is explained by an average inter-ACK time of $10.4$ ms for ACP+ in comparison to a much smaller $1.16$ ms for BBR-d1m1 that results from BBR-d1m1 attempting to achieve high throughputs. 

To summarize, ACP+ results in an average age of $115.5$ ms, an average delay of $110.79$ ms, an average throughput of $0.77$ Mbps and an inter-ACK time of $10.4$ ms. The corresponding values for BBR-d1m1 are $114.5$ ms, $112.33$ ms, $110$ Mbps and $1.16$ ms. \emph{ACP+ achieves an almost similar age as BBR-d1m1, however, at a significantly lower throughput.} The similar age at a much larger inter-ACK time is explained by the fact (observed in our experiments) that while a very low or high rate of updates results in high age, age stays relatively flat in response to a large range of update rates in between.  It turns out that ACP+ tends to settle in the flat region closer to where increasing the rate of updates stops reducing age. This much reduced throughput of ACP+ is especially significant in the context of shared access, allowing a larger number of end-to-end ACP+ flows to share an access without it becoming a bottleneck, as we had observed in Section~\ref{sec:orbit_results}. 

\subsubsection{The BBR Puzzle} 
What could explain the low age achieved by BBR-d1m1? We observe that the average delay of $112.33$ ms when using BBR-d1m1 is the same as that obtained by a \emph{Lazy} (introduced in~\cite{tanya-kaul-yates-wowmom2019}) status updating protocol we ran alongside the others, which sends an update once every round-trip time. One would expect \emph{Lazy} to achieve a round-trip time of $RTT_{\text{base}}$ (see Figure~\ref{fig:transient_tcp}). This tells us that BBR-d1m1's flow on an average saw an RTT of $RTT_{\text{base}}$. It obtained a low throughput of $100$ Mbps, which was an accidental consequence of the receiver buffer size settings of BBR-d1m1 that disallowed the congestion control algorithm to push bytes into the network at a larger rate. The higher throughput achieved by BBR-d1m3, as observed earlier, did come with a higher age.

\subsection{Age over Shared and Contended Access}
\label{sec:awsAccess}
We compare the performance of the congestion control algorithms with ACP+ for when one or more TCP clients connect to a server on the cloud via a WiFi access point. The setup and methodology are same as in Section~\ref{sec:orbit_results}. We use \texttt{iPerf3} to generate TCP traffic from ORBIT nodes towards the AWS server. We will first look at the \textit{low contention} configuration that has five or fewer nodes connect to the access point and then when there is \textit{high contention} and more than $10$ nodes share the WiFi access.

\subsubsection{Shared Network with Low Contention}
\label{sec:lowContention}
%
%
%
\begin{figure}[!t]             
	\begin{center}
		\subfloat[Average Age (ms)]{\includegraphics[width=.5\linewidth]{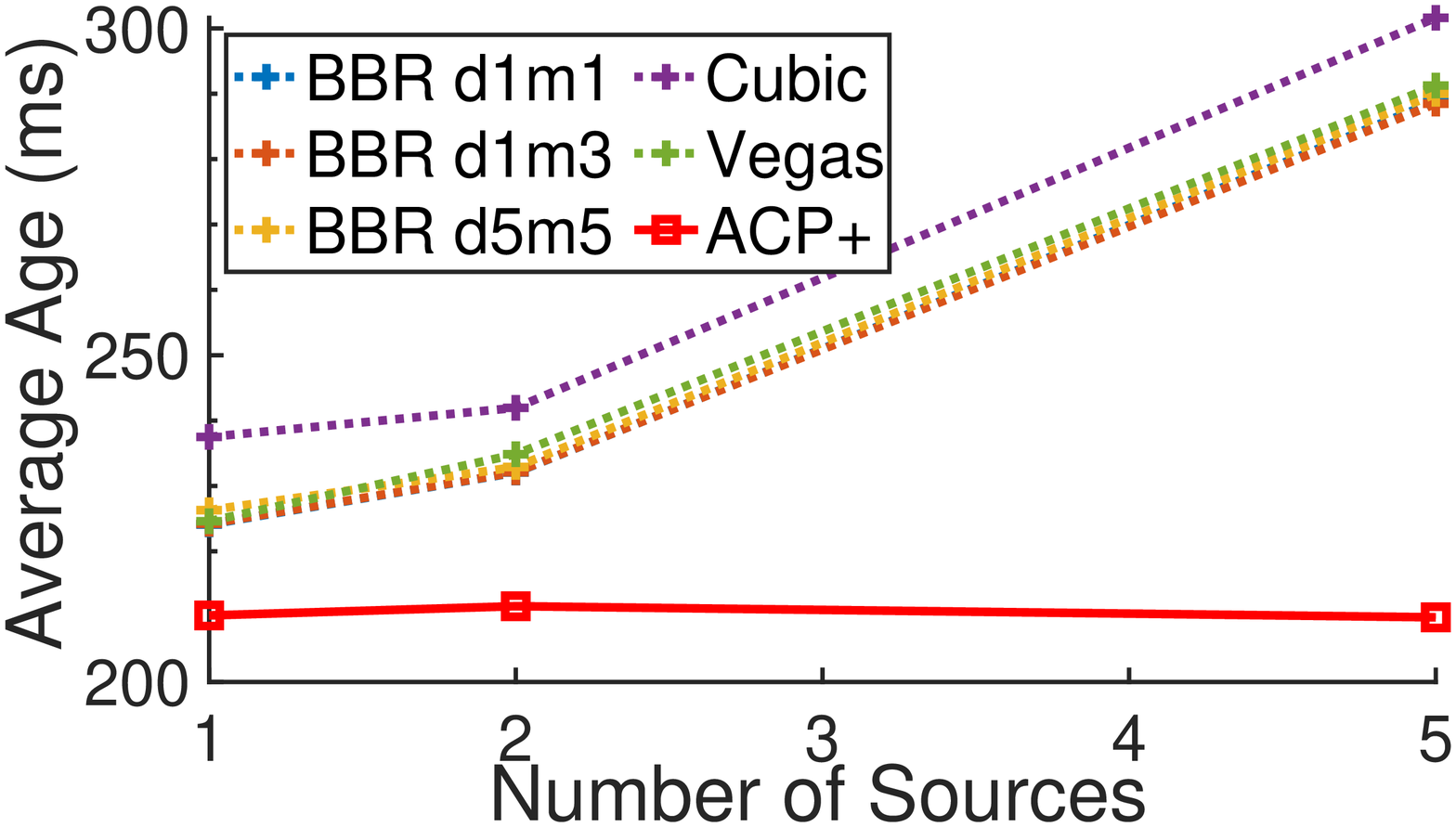}
			\label{fig:age_lowContention}}
		\subfloat[Sum Throughput (Mbps)]{\includegraphics[width=.5\linewidth]{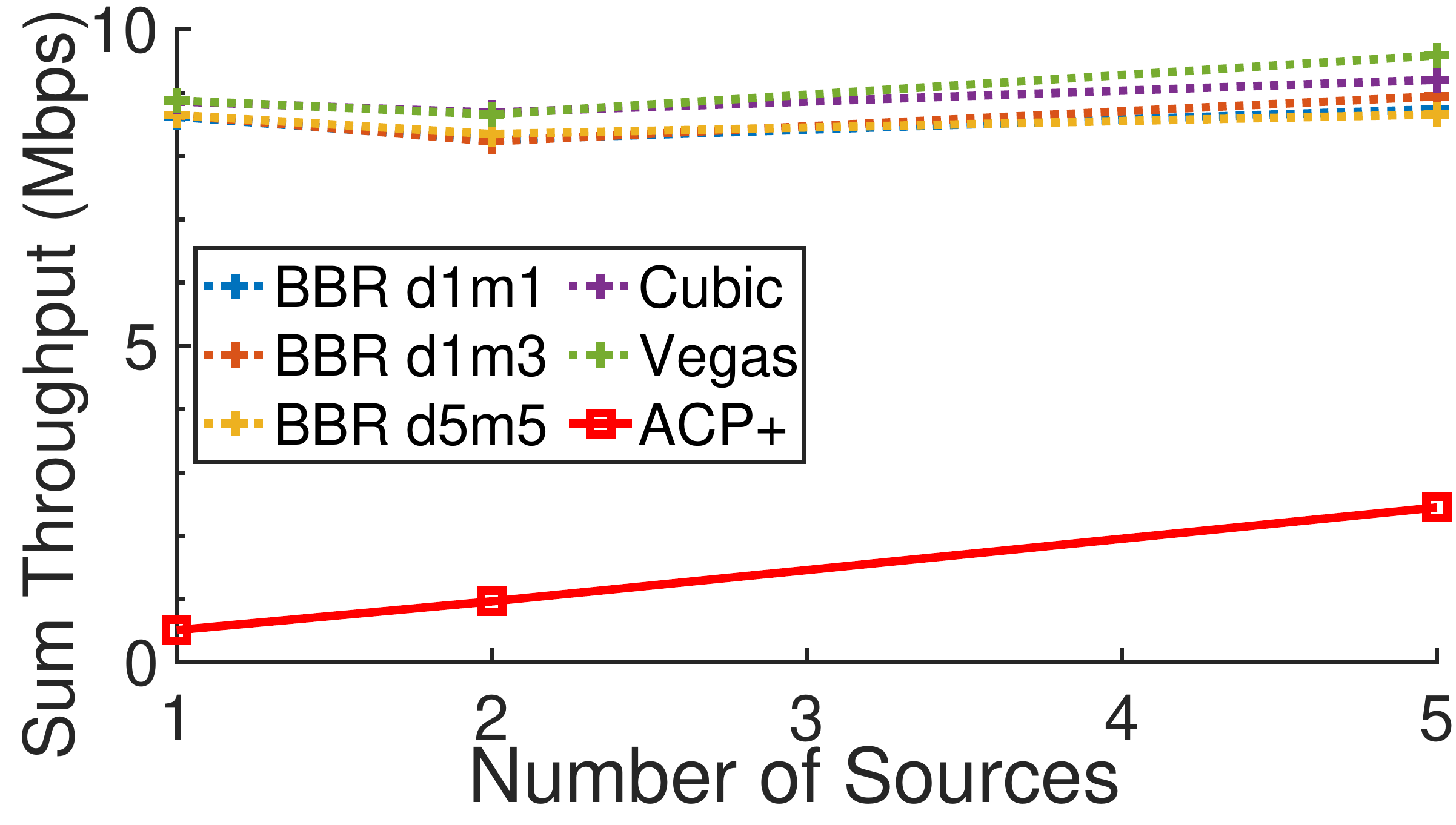}
			\label{fig:thr_lowContention}}
		\caption{Comparison of different TCP Congestion Control Algorithms and ACP+ under low WiFi access contention. The WiFi link rate was set to $12$ Mbps.}
		\label{fig:lowContention}
	\end{center}
\end{figure}

Figure~\ref{fig:lowContention} compares the different TCP congestion control algorithms and ACP+ in a low WiFi access contention environment. The number of nodes connected to the access point are a maximum of $5$. The WiFi link rate is set to $12$ Mbps. Figure \ref{fig:age_lowContention} shows the average age achieved by the TCP control algorithms and ACP+. ACP+ performs better than all the chosen congestion control algorithms and the gap between ACP+ and the rest increases as the number of clients increase from $1$ to $5$. 
%

The TCP algorithms, unlike ACP+, always fully utilize the bottleneck WiFi link. As is seen in Figure~\ref{fig:thr_lowContention}, TCP always achieves a sum throughput of about $8 - 9$ Mbps, which is close to saturating the $12$ Mbps WiFi link when we also include packet header overheads. 
Recall our results from Section~\ref{sec:orbit_results} for ACP+. The per source throughput at which age is minimized is very low. For a small enough number of sources the different ACP flows are oblivious to each other. The sum throughput, shown in Figure~\ref{fig:thr_lowContention}, is far from saturating the $12$ Mbps WiFi link.

\subsubsection{Shared Network with High Contention}\label{sec:highContention}

\begin{figure}[t]             
	\begin{center}
	        \subfloat[BBR Age]{\includegraphics[width=.5\linewidth]{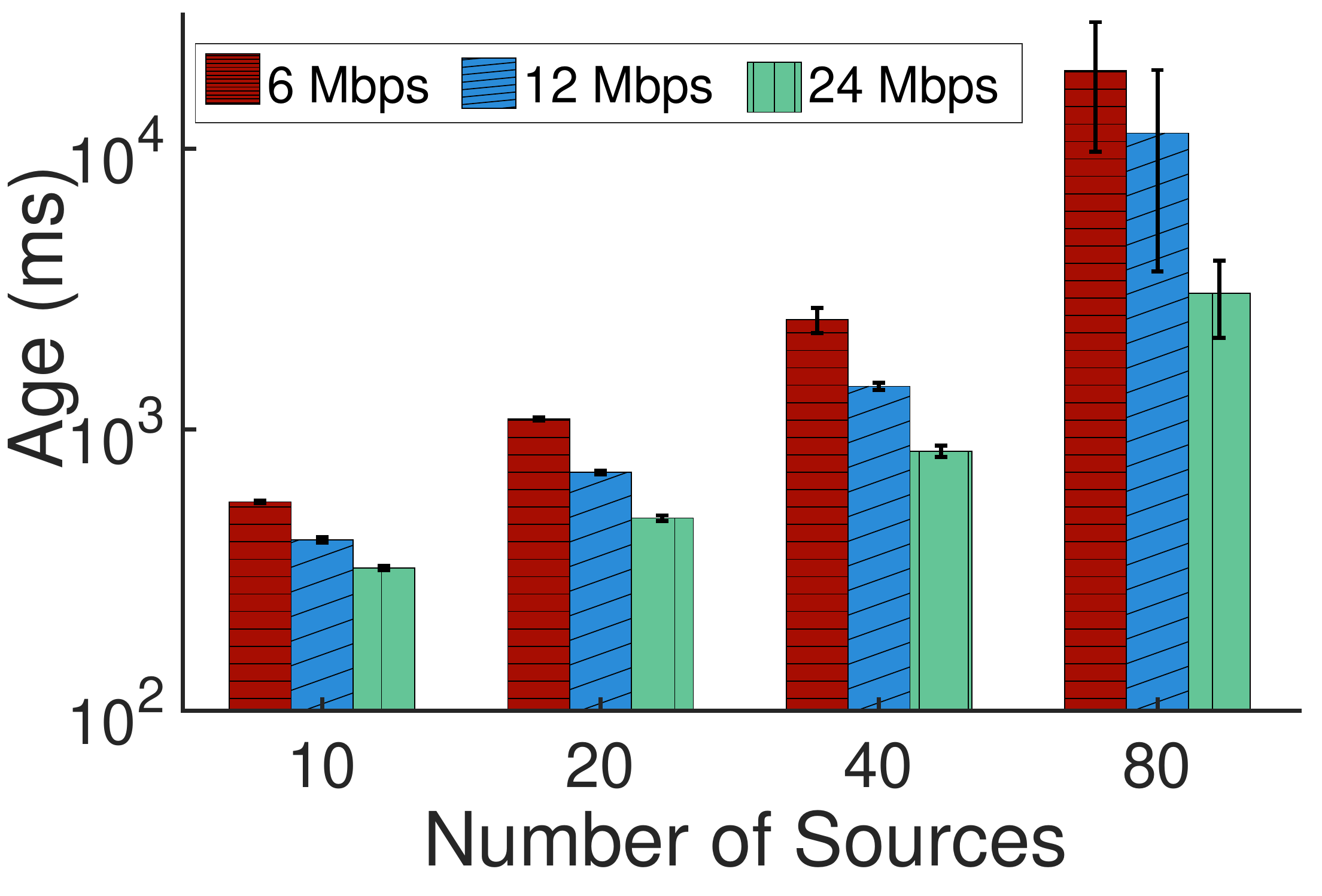}
			\label{fig:Age_BBR_HighDensity_bar}}
                \subfloat[BBR Throughput]{\includegraphics[width=0.5\linewidth]{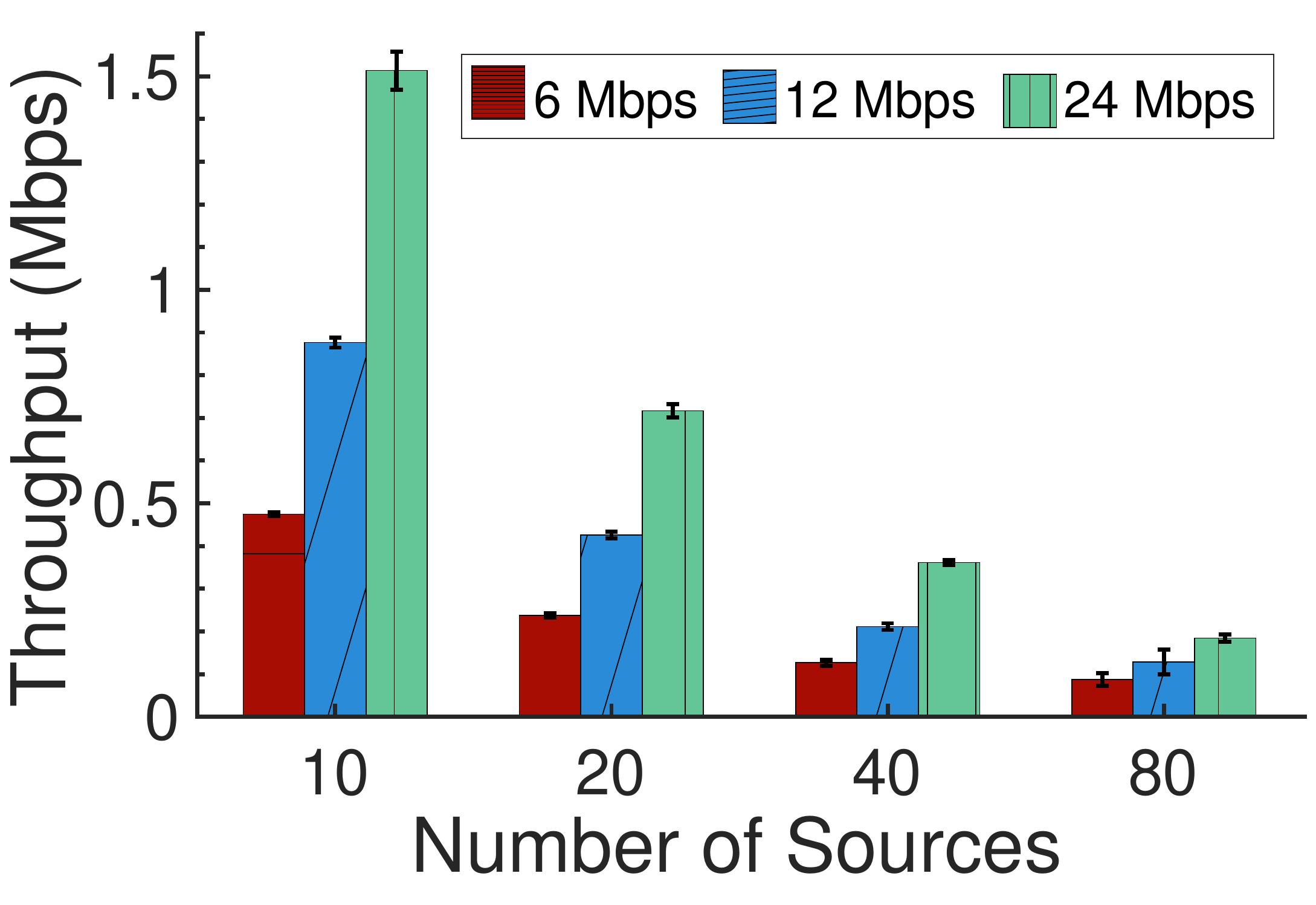}
			\label{fig:Thr_barTCP_ageing_high_density}}
		\caption{Average Age and per-node throughput achieved by BBR in the presence of high WiFi access contention. The solid region indicates the mean, and the bar indicates the standard deviation across different runs.}
		\label{fig:TCP_HighDensity_age_thr_bar}
	\end{center}
\vspace{-0.2in}
\end{figure}

In this network configuration, we only compare ACP+ and BBR since BBR outperforms CUBIC and Vegas in the core network (Section~\ref{sec:awsCore}) and also when WiFi access contention is low, as seen above. Figure~\ref{fig:Thr_barTCP_ageing_high_density} shows the age acheived by BBR for $10$, $20$, $40$ and $80$ clients connected to a fixed-rate WiFi access point, for rates $6$, $12$, and $24$ Mbps. Observe that, as we increase the number of clients for a given WiFi link rate, we see a very rapid increase in the average age per node achieved by BBR in comparison to the increase seen by sources using ACP+ (see Figure~\ref{fig:age_highContention}).

Figure~\ref{fig:Thr_barTCP_ageing_high_density} shows the BBR throughput per node. While BBR has larger per node throughputs than ACP+ (see Figure~\ref{fig:thr_highContention}), the throughputs for when there are $20$ or more clients are similar. For when there are $10$ clients, ACP+ has much smaller throughputs for WiFi link rates of $12$ and $24$ Mbps. This is because the ACP+ paths do not together fully utilize the WiFi link for the rates, since the constraining factor as regards optimization of age is the backhaul beyond the access (see Section~\ref{sec:orbit_results}). For $10$ clients at $6$ Mbps and for when there are $20$ or more clients, ACP+ and BBR have similar throughputs. However BBR achieves the throughputs at a much larger RTT (not shown) and hence achieves a much larger age. The large RTT when using BBR are likely because all BBR clients attempt to fully utilize the bottleneck link, which is the WiFi link in our experiments. ACP+, on the other hand, keeps the backlog in the end-to-end path small.

\section{Conclusion}\label{sec:conclusion}
We propose the \emph{Age Control Protocol} (ACP+) for applications that desire freshness of information at a destination (monitor), which is the first transport layer protocol that enables freshness of information at a destination, wherein the information is sent by a source in the form of status update packets to the destination over the Internet. We evaluate ACP+ over a wide range of simulated networks and real-world, end-to-end paths over the Internet. Our evaluation brings insights into age control over the Internet. We also compared and contrasted ACP+ with various state-of-the-art TCP congestion control algorithms. We also draw insights on why the existing algorithms are unsuitable for age control over the Internet. We make ACP+ publicly available at~\cite{acp-github}.

\ifCLASSOPTIONcaptionsoff
  \newpage
\fi

\bibliographystyle{IEEEtran} 
\bibliography{main}

\appendix

\section*{Analysis of Two Queues in Tandem}
\label{sec:twoQTandem}
We consider two queues, indexed $1,2$, in series. Queue $k$ is serviced by a server with exponentially distributed service times with mean $1/\mu_k$. Both queues have an infinitely large waiting room and in each queue service progresses in a FCFS manner. Status update packets arrive at queue $1$ as a Poisson process of rate $\lambda$. A status update packet enters queue $2$ on completion of service by server $1$. At the end of its service by server $2$, the packet is received by a monitor. 

Note that packets arrive into queue $2$ as a Poisson process of rate $\lambda$. We will assume that a packet that arrives into queue $2$ undergoes service for a time that is independent of the time it spent in service in server $1$. That is, somehow, correlations that may be introduced due to packet lengths do not exist and both the queues can be analyzed as $M/M/1/\infty$ queues.

The monitor tracks the age $\age(t)$ of the updates. In the following, we calculate the age-of-information (AoI) $\age = \lim_{t \to \infty} E[\age(t)]$ at the monitor. We have~\cite{KaulYatesGruteser-Infocom2012} 
\begin{align}
    \age = \lambda\left(\frac{E[X^2]}{2} + E[XT]\right),
\end{align}
where $X$ is the inter-arrival time between updates that arrive into the two queue system and $T$ is the system time, which is the total time that elapses between arrival of an update and its departure from queue $2$.
\subsection{Time-Average Analysis}
Let $T_i$ be the system time of packet $i$, which is the time elapsed between the arrival of packet $i$ into queue $1$ and its delivery to the monitor. Let $\Wait{i}{k}$ be the time spent by packet $i$ waiting in queue $k$ and let $\Service{i}{k}$ be the time spent by the packet in server $k$. The corresponding system time is $\System{i}{k} = \Wait{i}{k} + \Service{i}{k}$. Let $X_i$ be the time elapsed between the arrival of packet $i$ and packet $i-1$ to queue $1$. To calculate AoI, we must calculate $E[\System{i}{} X_i] = E[X_i E[\System{i}{}|X_i]]$. Further, 
\begin{align}
E[\System{i}{}|X_i] &= E[\Wait{i}{1}|X_i] + E[\Wait{i}{2}|X_i] + E[\Service{i}{1}] + E[\Service{i}{2}].
\label{eqn:ETiGivenXi}
\end{align}

We will next elaborate on the calculation of $E[\Wait{i}{2}|X_i]$. We will start by calculating $E[\Wait{i}{2} | \System{i}{1}, X_i]$. Note that the wait time $\Wait{i}{2}$ can be written as 
\begin{align}
\Wait{i}{2} = (\System{i-1}{} - \System{i}{1} - X_i)^{+}.
\end{align}

We want to calculate
\begin{align}
E[\Wait{i}{2} | \System{i}{1}, X_i] = E[(\System{i-1}{} - \System{i}{1} - X_i)^{+}| \System{i}{1}, X_i].
\label{eqn:Wi2GivenTi1Xi}
\end{align}
To do so, we will calculate the conditional pdf $f_{\System{i-1}{}|\System{i}{1},X_i}(.)$. This pdf may be calculated as the convolution of the pdfs $f_{\System{(i-1)}{1}|\System{i}{1},X_i}(.)$ and $f_{\System{(i-1)}{2}}$. This is because the random variable $\System{i-1}{} = \System{(i-1)}{1} + \System{(i-1)}{2}$ and $\System{(i-1)}{2}$ is independent of $\System{i}{1}$ and $X_i$. The marginal pdf $f_{\System{(i-1)}{2}}$ is simply the steady state distribution of system time in a $M/M/1/\infty$ queue with arrivals at rate $\lambda$ and service at rate $\mu_2$. We have
\begin{align}
f_{\System{(i-1)}{2}}(t) = 
\begin{cases}
(\mu_2 - \lambda) e^{-(\mu_2 - \lambda)t} & t\ge 0,\\
0 & \text{otherwise}.
\end{cases}
\label{eqn:Ti_12}
\end{align}
We can rewrite the pdf $f_{\System{(i-1)}{1}|\System{i}{1},X_i}$ as
\begin{align}
f_{\System{(i-1)}{1}|\System{i}{1},X_i} = \frac{f_{\System{i}{1}|\System{(i-1)}{1},X_i}\,  f_{\System{(i-1)}{1}}}{f_{\System{i}{1}|X_i}},
\label{eqn:Ti_1GivenTiXi}
\end{align}
where we used the chain rule and the fact that $\System{(i-1)}{1}$ is independent of $X_i$.

To calculate the pdf $f_{\System{i}{1}|\System{(i-1)}{1},X_i}$, observe that $\System{i}{1} = (\System{(i-1)}{1} - X_i)^{+} + \Service{i}{1}$. We have
\begin{align}
f_{\System{i}{1}|\System{(i-1)}{1},X_i}(t|u,x) =
\quad \begin{cases}
\mu_1 e^{-\mu_1 (t - (u - x)^{+})} & t\ge (u - x)^{+},\\
0 & \text{otherwise}.
\label{eqn:TiGivenTi_1Xi}
\end{cases}
\end{align}
Let $\gamma_k = \mu_k - \lambda$. The pdf $f_{\System{(i-1)}{1}}$ is the steady state distribution of the system time for a $M/M/1$ queue with arrivals at rate $\lambda$ and service at rate $\mu_1$. 
\begin{align}
f_{\System{(i-1)}{1}}(u) =
\begin{cases}
\gamma_1 e^{-\gamma_1 u} & u\ge 0,\\
0 & \text{otherwise}.
\label{eqn:Ti_1}
\end{cases}
\end{align}
The pdf $f_{\System{i}{1}|X_i}$ is the conditional pdf of system time of packet $i$ conditioned on the inter-arrival time between packet $i$ and $i-1$ for a $M/M/1$ queue with arrivals at rate $\lambda$ and service at rate $\mu_1$.
\begin{align}
&f_{\System{i}{1}|X_i}(u|x) =\nonumber\\
&\begin{cases}
\mu_1 (1 - e^{-\gamma_1 x})  e^{-\mu_1 u} - \frac{\mu_1 \gamma_1}{\lambda} (e^{-\mu_1 u} - e^{-\gamma_1 u}) e^{-\gamma_1 x} & u\ge 0,\\
0 & \text{else}.
\end{cases}
\label{eqn:Ti_1GivenX_i}
\end{align}

Substituting~(\ref{eqn:TiGivenTi_1Xi}),~(\ref{eqn:Ti_1}) and~(\ref{eqn:Ti_1GivenX_i}), in~(\ref{eqn:Ti_1GivenTiXi}) gives us the pdf $f_{\System{(i-1)}{1}|\System{i}{1},X_i}$. Convolving with the pdf~(\ref{eqn:Ti_12}) we get $f_{\System{i-1}{}|\System{i}{1},X_i}(.)$. We use this pdf to calculate the conditional expectation in~(\ref{eqn:Wi2GivenTi1Xi}). Using the pdf~(\ref{eqn:Ti_1GivenX_i}) we can then calculate $E[(\System{(i-1)}{} - \System{i}{1} - X_i)^+|X_i]$ followed by $E[X_i(\System{(i-1)}{} - \System{i}{1} - X_i)^+] = E[\Wait{i}{2}X_i]$. Next we list the expectations that enable us to calculate $E[\System{i}{} X_i]$.
\begin{align}
&E[\Wait{i}{2} X_i] = \frac{\lambda}{\mu_2^2 (\mu_2 - \lambda)} + \frac{\lambda^2}{\mu_1 \mu_2 (\mu_1 + \mu_2 - \lambda)},\nonumber\\
&E[\Wait{i}{1} X_i] = \frac{\lambda}{\mu_1^2 (\mu_1 - \lambda)},\, 
E[\Service{i}{k} X_i] = \frac{1}{\lambda \mu_k}.\nonumber
\end{align}
In steady-state, for all packets $i$, $W_i =^{\text{st}} W$, $T_i =^{\text{st}} T$, and $X_i =^{\text{st}} X$. The AoI $\age$ is
\begin{align}
\age &= \frac{1}{\lambda} + \frac{1}{\mu_1} + \frac{1}{\mu_2} + \frac{\lambda^2}{\mu_1^2 (\mu_1 - \lambda)} +\nonumber\\ &\qquad \frac{\lambda}{\mu_2^2 (\mu_2 - \lambda)} + \frac{\lambda^2}{\mu_1 \mu_2 (\mu_1 + \mu_2 - \lambda)}.
\end{align}

\end{document}